\documentclass[twocolumn,aps,prd,showpacs,nofootinbib,superscriptaddress,preprintnumbers,floatfix,10pt]{revtex4-1}

\usepackage{amsmath}
\usepackage{amsfonts}
\usepackage{amssymb}
\usepackage{epsfig}
\usepackage{graphicx}
\usepackage{color}
\usepackage{slashed}
\usepackage{epstopdf}

\usepackage{hyperref}

\newcommand{\I}{\mathrm{i}}
\newcommand{\E}{\mathrm{e}}

\newcommand{\STr}{\text{STr}}

\newcommand{\Nc}{N_{\mathrm{c}}}
\newcommand{\Nf}{N_{\mathrm{f}}}
\newcommand{\pt}{\partial_t}
\newcommand{\psib}{\bar{\psi}}

\newcommand{\trho}{\tilde{\rho}}

\newcommand{\UL}{U_{\Lambda}}

\newcommand{\mL}{m_\Lambda^2}
\newcommand{\lL}{\lambda_{2,\Lambda}}
\newcommand{\mH}{m_{\text{H}}}
\newcommand{\mtop}{m_{\text{t}}}

\newcommand{\Nu}{N_{\text{u}}}
\newcommand{\Nh}{N_{\mathrm{h}}}

\newcommand{\rmB}{\mathrm{B}}
\newcommand{\rmF}{\mathrm{F}}
\newcommand{\rmG}{\mathrm{G}}
\newcommand{\muphi}{\mu_{\phi}^{2}}
\newcommand{\mutop}{\mu_{\mathrm{t}}^{2}}

\newcommand{\HL}{H_\Lambda}

\newcommand{\UMF}{U^{\mathrm{MF}}}
\newcommand{\HMF}{H^{\mathrm{MF}}}
\newcommand{\UEMF}{U^{\mathrm{EMF}}}
\newcommand{\HEMF}{H^{\mathrm{EMF}}}
\newcommand{\SU}{\mathrm{SU}}
\newcommand{\rmT}{\mathrm{T}}
\newcommand{\eff}{\mathrm{eff}}

\begin{document}

\preprint{}

\title {Impact of generalized Yukawa interactions on the lower Higgs mass bound}

\author{Holger Gies}
\email{holger.gies@uni-jena.de}
\affiliation{Theoretisch-Physikalisches Institut, Friedrich-Schiller-Universit\"at Jena, Max-Wien-Platz 1, D-07743 Jena, Germany}
\affiliation{Abbe Center of Photonics, Friedrich-Schiller-Universit\"at Jena, Max-Wien-Platz 1, D-07743 Jena, Germany}
\affiliation{Helmholtz-Institut Jena, Fr\"obelstieg 3, D-07743 Jena, Germany}
\author{Ren\'{e} Sondenheimer}
\email{rene.sondenheimer@uni-jena.de}
\affiliation{Theoretisch-Physikalisches Institut, Friedrich-Schiller-Universit\"at Jena, Max-Wien-Platz 1, D-07743 Jena, Germany}
\author{Matthias Warschinke}
\email{m\_warschinke@chiba-u.jp}
\affiliation{Theoretisch-Physikalisches Institut, Friedrich-Schiller-Universit\"at Jena, Max-Wien-Platz 1, D-07743 Jena, Germany}
\affiliation{Department of Physics, Graduate School of Science, Chiba University,  
Chiba 263-8522, Japan}

\begin{abstract}
  We investigate the impact of operators of higher canonical dimension
  on the lower Higgs mass consistency bound by means of generalized
  Higgs--Yukawa interactions. Analogously to higher-order operators in
  the bare Higgs potential in an effective field theory approach, the
  inclusion of higher-order Yukawa interactions, e.g.,
  $\phi^3\psib\psi$, leads to a diminishing of the lower Higgs mass
  bound and thus to a shift of the scale of new physics towards larger
  scales by a few orders of magnitude without introducing a
  metastability in the effective Higgs potential.  We observe that similar
  renormalization group mechanisms near the weak-coupling fixed point
  are at work in both generalizations of the microscopic action.  Thus, a combination of higher-dimensional operators
  with generalized Higgs as well as Yukawa interactions does not lead
  to an additive shift of the lower mass bound, but relaxes the
  consistency bounds found recently only slightly. On the method side,
  we clarify the convergence properties of different projection and
  expansion schemes for the Yukawa potential used in the functional
  renormalization group literature so far.
\end{abstract}


\maketitle

\section{Introduction}
\label{intro}

The discovery of the Higgs boson at the LHC \cite{Aad:2012tfa,
  Chatrchyan:2012xdj} was the major success of LHC run I.
Nonetheless, the result of the accurate measurement of the Higgs mass
of $125.09$ GeV \cite{Aad:2015zhl} is puzzling in view of mass bounds
predicted within the standard model
\cite{Krive:1976sg,Krasnikov:1978pu,Hung:1979dn,Politzer:1978ic,Linde:1979ny,Lindner:1988ww,Sher:1988mj,Arnold:1989cb,Arnold:1991cv,Ford:1992mv,Sher:1993mf,Altarelli:1994rb,Espinosa:1995se,Isidori:2001bm,Einhorn:2007rv,Ellis:2009tp}.
Though the measured value lies close to the lower mass bound, it seems
to slightly violate it.  This near criticality has lead to an ongoing
debate about the stability of the electroweak vacuum and the life time
of our Universe within the standard model
\cite{Alekhin:2012py,Gabrielli:2013hma,Buttazzo:2013uya,Bednyakov:2015sca,DiLuzio:2015iua}
as well as in various models beyond
\cite{Espinosa:1991gr,Casas:1994qy,Casas:1996aq,Bergerhoff:1999jj,Espinosa:2007qp,ArkaniHamed:2008ym,EliasMiro:2011aa,Kadastik:2011aa,EliasMiro:2012ay,Anchordoqui:2012fq,Lebedev:2012zw,Chen:2012faa,Masina:2012tz,Bhattacharjee:2012my,Krajewski:2014vea,Haba:2014qca,Das:2015nwk,Loebbert:2015eea,Czerwinska:2015xwa,Espinosa:2015zoa,Abe:2016irv,Ema:2016ehh}.

The usual perturbative treatment of the lower Higgs mass bound is
based on the assumption that the high energy behavior of particle
physics is dominated only by standard-model degrees of freedom as well
as by perturbatively renormalizable operators. The latter is well
justified in the case that nature is close to the Gau\ss ian fixed point
during the entire renormalization group (RG) flow. Hence, the
influence of higher-order operators on infrared (IR) observables is
highly suppressed due to power counting arguments and thus far beyond
any experimental tests.
In this spirit, the stability of the electroweak vacuum is only connected to the sign of the running quartic Higgs self-coupling \cite{Buttazzo:2013uya,Bednyakov:2015sca,Iacobellis:2016eof}.

However, even if the impact of higher-dimensional operators is
negligibly small in the IR, their presence may influence the RG
running of the relevant and marginal couplings in the deep ultraviolet
(UV).  From a Wilsonian RG point of view, it is, in fact, natural to
expect that the dimensionless couplings of these operators are of
order $\mathcal{O}(1)$ at the UV scale, below which a description in
terms of the standard model degrees of freedom becomes meaningful.
Indeed, recent studies based on general RG arguments in an effective
field theory approach demonstrate that higher-dimensional operators at
some UV cutoff scale $\Lambda$ can exert an influence on the
Higgs-mass stability bound.

In these studies, based on the functional RG
\cite{Gies:2013fua,Gies:2014xha,Jakovac:2015kka,Jakovac:2015iqa} as
well as nonperturbative lattice simulations
\cite{Holland:2003jr,Holland:2004sd,Gerhold:2007yb,Gerhold:2007gx,Gerhold:2009ub},
a lower Higgs mass bound emerges naturally from the RG flow itself and
is primarily connected with a consistency condition on the bare action
rather than with the stability of the effective Higgs potential which
is a resulting long-range property.  Moreover, it can be shown that
higher-order operators at the standard-model cutoff scale can have a
quantitative impact on the lower Higgs mass bound as well as the
stability properties of the bare potential. For instance, a positive
$\phi^6$ operator at the Planck scale leads to a diminishing of the lower Higgs mass bound
by $\sim 1$ GeV introducing neither a
metastability nor an instability during the entire RG flow from the
cutoff scale $\Lambda$ to the deep IR of the scalar potential
\cite{Eichhorn:2015kea}. For cutoff scales below the Planck scale, the
effect on the Higgs mass bounds is even larger. Of course, these
studies do not imply that metastabilities can be excluded. In fact, a
new lower Higgs mass bound in the presence of higher-order Higgs
self-couplings might be defined by the requirement of absolute
stability at all scales. Demanding even lower Higgs masses leads to
metastable effective potentials in which the metastability is seeded
in the bare potential and thus by the underlying theory of the
standard model \cite{Borchardt:2016xju}. For an example of such a
scenario, see \cite{Hebecker:2013lha}.

The aim of this work is to investigate the impact of further
higher-dimensional operators by means of a generalized Yukawa
interaction term. This goes beyond the class of polynomially stable
bare potentials (e.g., of $\phi^6$-type) with a unique minimum studied
so far. The latter facilitate to lower conventional Higgs mass bounds
by a limited amount when requiring stable scalar potentials during the
entire RG flow in the weak-coupling domain. The inclusion of mixed
fermion-scalar operators allows to pursue also another direction.  The
lower Higgs mass bound within $\phi^4$ bare potentials is mainly
dominated by the Yukawa sector, i.e., the large top Yukawa coupling.
Thus, the lower bound can be shifted towards lighter Higgs masses if
the values of the running Yukawa coupling are smaller during the RG
flow.  Any mechanism that lowers the value of the bare top Yukawa
coupling can also lower the resulting Higgs mass. In the same way as
generalized bare potentials allow smaller values for the bare quartic
coupling in order to lower the bound, a generalized bare Yukawa
potential could lead to smaller values of the bare Yukawa coupling
with still a phenomenologically correct IR description of the
top mass. Exploring this new mechanism is the goal of this work.

This paper is organized as follows: in Sec. \ref{sec:model} we introduce our toy models and present RG flow equations at an extended mean-field as well as a nonperturbative level. Section \ref{sec:projection} discusses different projection rules on the Yukawa functional as they are used in the literature and clarifies which projection rule leads to the fastest convergence. In Sec. \ref{sec:pheno} we concentrate on the phenomenological consequences for the lower Higgs mass bound if we allow for generalized Yukawa interactions at some UV cutoff scale of the theory.
Our main results are summarized in Sec. \ref{sec:conclusions}.

\section{RG flow of Higgs-Yukawa models}
\label{sec:model}
All relevant points, which we would like to address in this work, can be illustrated in the following (gauged) Higgs--Yukawa toy model.
The model consists of a real scalar field $\phi$ and a Dirac fermion $\psi$. Both fields are coupled via a simple Yukawa interaction term, $\phi\psib\psi$. This reduces drastically the complex chiral structure of the Higgs-fermion sector of the standard model, however, this reduction is justified by the fact that only the top quark has a major impact due to the negligible Yukawa couplings of the other fermions. Moreover, we neglect the remaining electroweak sector, i.e., the electroweak gauge bosons. As the electroweak sector is a gauge theory, every observable has to be formulated in a strictly gauge invariant manner. 
Fortunately, the Fr\"ohlich-Morchio-Strocchi mechanism \cite{Frohlich:1981yi,Frohlich:1980gj} maps properties of the gauge-invariant composite states to properties of the elementary fields in the Lagrangian which has been demonstrated in the standard model by nonperturbative lattice simulations in the pure gauge-Higgs sector \cite{Maas:2012tj,Maas:2013aia,Maas:2014pba} but can be done also for the fermions \cite{Egger:2017tkd}. 
For theories beyond the standard model these type of mechanisms have recently attracted interest \cite{Maas:2015gma,Maas:2016qpu,Maas:2016ngo}.
Neglecting the electroweak gauge part, we avoid these intricate subtleties from the nontrivial gauge-Higgs interplay.
The main advantage of this reduction is that we can concentrate on the considered mechanisms.

However, there is a significant difference between the RG flow of the Yukawa coupling in a simple Yukawa model in comparison with the full standard model of particle physics. The running Yukawa coupling decreases towards the UV in the standard model due to gauge boson contributions. By contrast, gauge boson fluctuations are absent in a simple Higgs-Yukawa toy model and the Yukawa coupling perturbatively runs into a pole for large RG scales. In order to study possible phenomenological influences on the Higgs mass bounds via the different runnings of generalized Yukawa interactions, we also investigate a toy model in which the top quark is gauged under the $\SU(3)$ symmetry group of the strong interaction. In the gauged version, the Yukawa coupling is asymptotically free.

The Euclidean action of the model reads 
\begin{align}
\begin{split}
 S_{\mathrm{cl}} = \int_x \bigg[ &\frac{1}{2}(\partial_\mu\hat{\phi})^2 + U(\hat{\rho}) + \psib^a \I \slashed{D}^{ab} \psi^b + \frac{1}{4}(F_i^{\mu\nu})^2 \\
 & + \I H(\hat{\rho}) \hat{\phi} \psib^a\psi^a  \bigg].
\end{split}
\label{eq:ClassicalAction}
\end{align}
The hat indicates that $\hat{\phi}$ is an unrenormalized bare scalar field. Renormalized quantities, such as the renormalized scalar field $\phi$, are introduced below.
We impose a discrete $\mathbb{Z}_2$ symmetry on the scalar potential, $\hat\rho = \hat{\phi}^{2}/2$, in order to mimic the structure of the electroweak symmetry group in the scalar sector. The symmetry extends to the full action being invariant under a discrete chiral symmetry, $\hat\phi \to -\hat\phi$, $\psi \to \E^{\I \frac{\pi}{2}\gamma_*}\psi$, and $\psib \to \psib \E^{\I\frac{\pi}{2}\gamma_*}$. Thus, a mass term for the fermions is forbidden and has to be generated through spontaneous symmetry breaking. In order to study the impact of higher-dimensional operators on the Higgs mass bound, we introduce a generic Yukawa potential $H(\hat{\rho})$, defining a separate Yukawa coupling for every field amplitude $\hat\phi$. The fermion is coupled to the Gluon $G^{\mu}_i$ via the standard covariant derivative in the fundamental representation, $D_{\mu}^{ab} = \delta^{ab}\partial_\mu + \I \bar{g} G_{i\,\mu} \, T_i^{ab}$, $F_i^{\mu\nu}$ is the field strength tensor, and $\bar{g}$ the unrenormalized gauge coupling of the strong force.

\subsection{Nonperturbative RG equations}
\label{sec:flow}
In order to derive flow equations not only for separate couplings but rather for the full Yukawa and scalar potential, we use the functional RG. This has several advantages. 
First, the functional RG yields a closed-form flow equation for every higher-order operator present in both potentials, $H(\hat{\rho})$ and $U(\hat{\rho})$. More precisely, these flow equations are given by beta functionals, $\beta_H(\hat{\rho})$ and $\beta_U(\hat{\rho})$, defining a beta function for the potentials at any field value $\hat\phi$. The running of particular (higher-dimensional) couplings can be extracted as expansion coefficients with respect to the scalar field from these closed-form beta functionals. 
Second, the investigation of the full beta functionals allows to keep track of all relevant scales in the system, i.e., we are not limited to conventional approximation schemes like $\mu=\phi$ which mixes momentum scale information with the field amplitude.
Third, the functional RG includes resummations of certain classes of diagrams, allows for treating strong coupling regimes, and incorporates threshold effects as well.

The central object in the mathematical implementation of the functional RG is the effective average action $\Gamma_k$ which interpolates between the bare action, $\Gamma_{k=\Lambda}=S$, defining the theory at some microscopic (high-momentum) UV scale $\Lambda$ and the effective action, $\Gamma_{k=0}=\Gamma$, describing the resulting IR physics after all fluctuations are integrated out. The induced flow in theory space between these two actions is governed by the Wetterich equation \cite{Wetterich:1992yh},
\begin{align}
 \pt \Gamma_k = \frac{1}{2} \STr \frac{\pt R_k}{\Gamma^{(2)}_k + R_k^{}}.
 \label{eq:Wetterich}
\end{align}
The RG scale $k$ is introduced via an IR regulator $R_k(p)$ acting as an additional mass term for the full propagator and ensures IR finiteness. Moreover, the regulator function is chosen such that its derivative $\pt R_k$, with $\pt = \mathrm{d}/(\mathrm{d}\ln k)$ is peaked at momentum $p^2 \approx k^2$ in order to implement the Wilsonian concept of integrating out quantum fluctuations momentum shell by momentum shell. Furthermore, the numerator in Eq.~\eqref{eq:Wetterich} ensures that the flow does not suffer from UV divergencies. For detailed reviews on the functional RG see \cite{Morris:1998da,Bagnuls:2000ae,Berges:2000ew,Aoki:2000wm,Polonyi:2001se,Pawlowski:2005xe,Gies:2006wv,Delamotte:2007pf,Braun:2011pp,Rosten:2010vm}.

In order to solve the flow equation ($\ref{eq:Wetterich}$) for the considered gauged Higgs--Yukawa model, we have a choice between two possibilities of how to treat the gauge redundancy of the model. In principle, one can construct gauge-invariant flow equations which provide a conceptually clean set up \cite{Morris:1998kz,Branchina:2003ek,Pawlowski:2003sk,Arnone:2005vd,Arnone:2005fb,Morris:2006in,Wetterich:2016ewc}. However, in practice it is useful to work with a gauge-fixed formulation.
For the gauge fixing we adopt the conventional Lorenz gauge $\partial_{\mu}G_i^{\mu} = 0$ (or its background-gauge variant). Hence our ansatz for the gauge-fixed effective average action reads
\begin{align}
 \Gamma_k = \int_x \bigg[&
 \frac{Z_{\phi}}{2}(\partial_\mu \hat\phi)^2 + U(\hat\rho) + Z_{\psi}\psib^a \I \slashed{D}^{ab} \psi^b 
 \notag \\ &+ \I H(\hat\rho) \hat\phi \psib^a\psi^a 
 + \frac{Z_{\rmG}}{4} (F_i^{\mu\nu})^2 + \frac{Z_{\rmG}}{2\xi}(\partial_{\mu}G_i^{\mu})^2 \notag \\ 
 &+ Z_{c}\bar{c}_i \partial_{\mu}D^{\mu}_{ij}c_j
 \bigg],
 \label{eq:EffectiveAverageAction}
\end{align}
where $\xi$ is the gauge-fixing parameter, $D^\mu_{ij}$ the covariant derivative in the adjoint representation, and $c_i$ and $\bar c_i$ are the Faddeev--Popov ghosts.
Note, that this ansatz is based on a systematic derivative expansion. This truncation has proven useful for this class of (gauged or ungauged) models in various physical contexts, such as QCD \cite{Jungnickel:1995fp,Bohr:2000gp,Braun:2008pi,Pawlowski:2014zaa,Braun:2014ata}, solid state and statistical physics \cite{Rosa:2000ju,Hofling:2002hj,Braun:2010tt,Diehl:2007th,Classen:2015mar}, supersymmetry \cite{Synatschke:2008pv,Gies:2009az,Heilmann:2014iga,Hellwig:2015woa,Gies:2017tod}, or Higgs physics \cite{Gies:2013fua,Gies:2013pma,Eichhorn:2014qka,Gies:2015lia,Gies:2016kkk,Jakovac:2015kka}. The impact of higher-derivative terms can be estimated by comparing leading-order results, i.e., a local potential approximation $Z_{\phi,k} = 1 = Z_{\psi,k}$, with results including a running wave function renormalization. A breakdown of this expansion can be indicated by anomalous dimensions, $\eta = -\pt \ln Z$, growing large.

The nonperturbative flow equations for the scalar potential and the Yukawa potential as well as for the other quantities can be extracted by plugging the ansatz \eqref{eq:EffectiveAverageAction} into the Wetterich equation \eqref{eq:Wetterich}. 
It is useful and convenient to introduce renormalized fields, e.g., $\phi = Z_{\phi}^{1/2} \hat\phi$, to fix the usual RG invariance of field rescalings, as well as to introduce dimensionless renormalized quantities, such as
\begin{align}
\begin{split}
 \trho &= Z_{\phi} k^{2-d} \hat\rho, \qquad u(\trho) = k^{-d} U_k(\hat\rho), \\
 g^2 &= Z_{\rmG}^{-1} k^{4-d} \bar{g}^2, \quad
 h(\trho) = Z_{\phi}^{-\frac{1}{2}} Z_{\psi}^{-1} k^{\frac{4-d}{2}} H_k(\hat\rho).
\end{split}
\end{align}
We suppress the explicit dependence on the RG scale $k$ of the various
running quantities for better readability.  The flow of the
dimensionless scalar, $u_k$, and Yukawa potential, $h_k$, for fixed $\trho$ finally
reads:
\begin{align}
  \pt \, u =&  -d \, u + (d-2+\eta_{\phi})\trho u' \label{eq:flowPot} \\
            &+ 2 v_{d} \Big[ l_0^{(\rmB)d}(u' + 2 \trho u'';\eta_{\phi}) - d_{\gamma}\Nc \, l_{0}^{(\rmF)d}(2 \trho h^{2} ; \eta_{\psi}) \Big],
            \notag \\
 \begin{split}
  \pt \, h =&\, \frac{1}{2} (d-4 + \eta_\phi + 2\eta_\psi) h  +  (d-2 + \eta_\phi)\trho h' \\ 
            &+  2v_d \Big[2 h (h + 2\trho h')^2  l_{1,1}^{(\rmF\rmB)d}(2 \trho h^2,u' + 2\trho u'' ; \eta_\psi,\eta_\phi) \\ 
            &-  (3h' + 2\trho h'') l_{1}^{(\rmB)d}(u' + 2\trho u'';\eta_{\phi}) \\
            &- 2 g^2 h \frac{\Nc^2-1}{2\Nc}(d-1+\xi) \, l_{1,1}^{(\rmF\rmB)d}(2\trho h^2,0; \eta_\psi,\eta_\rmG) \Big].
 \end{split}
 \label{eq:flowYukawa}
\end{align}
The primes denote derivatives with respect to the dimensionless renormalized field invariant $\trho$, and $v_d^{-1} = 2^{d+1}\pi^{\frac{d}{2}}\Gamma(d/2)$. $d_{\gamma}$ denotes the dimension of the Clifford algebra. Throughout this work, we use $d_{\gamma}=4$ for all numerical calculations. Irrelevant field-independent contributions to the scalar potential are ignored.  The threshold functions $l_n^{(\rmB/\rmF)d}$ and $l_{1,1}^{(\rmF\rmB)d}$  arise from the integration over the loop momentum and entail the non-universal regulator dependence. Their general definitions as well as explicit representations for a convenient regulator choice are listed in the appendix. 
The nonperturbative running of the wave function renormalizations $Z_{\phi}$ and $Z_{\psi}$ can be encoded in the anomalous dimensions of the scalar and fermion field,
\begingroup
\allowdisplaybreaks{
\begin{align}
 \eta_{\phi}  =&  \frac{8v_d}{d} \Big[ \kappa \big(3 u''(\kappa) + 2\kappa u'''(\kappa)\big)^2 m_2^{(\rmB)d}(\muphi;\eta_{\phi}) \notag \\
               &\qquad + d_{\gamma}\Nc \big(h(\kappa)+2\kappa h'(\kappa)\big)^2 \Big( m_4^{(\rmF)d}(\mutop;\eta_{\psi}) \notag \\
               &\qquad - 2\kappa h^2(\kappa) \, m_2^{(\rmF)d}(\mutop;\eta_{\psi}) \Big) \Big],
 \\
 \eta_{\psi}  =&  \frac{8v_d}{d} \bigg[ (h(\kappa) + 2\kappa h'(\kappa))^2 m_{1,2}^{(\rmF\rmB)d}(\mutop,\muphi;\eta_\psi,\eta_\phi) \notag \\
               & -\frac{\Nc^2-1}{2\Nc} g^2  \Big[ (1-d+\xi)m_{1,2}^{(\rmF\rmB)d}(\mutop,0;\eta_\psi,\eta_\rmG) \notag \\
               &+ (1-d)(\xi-1)\tilde{m}_{1,1}^{(\rmF\rmB)d}(\mutop,0;\eta_\psi,\eta_\rmG) \Big] \bigg],
               \label{eq:anomalFermion}
\end{align}
}%
\endgroup%
where the threshold functions can be found in the appendix again, $\muphi = u'(\kappa)+2\kappa u''(\kappa)$ and $\mutop = 2\kappa h(\kappa)^{2}$, and $\kappa$ denotes the flowing minimum of the scalar potential. 
These flow equations reduce to the set of flow equations of the simple ungauged Higgs--Yukawa model by setting $g=0$ and $\Nc=1$. 

Flow equations for a field-dependent Yukawa coupling in the ungauged
model were already studied in the context of quark-meson models at
finite temperature \cite{Pawlowski:2014zaa} as well as in the Higgs
context \cite{Jakovac:2015kka} recently. A
  reparametrization in terms of a Yukawa function $\tilde{h}(\phi)$
  corresponding to $\tilde{h}(\phi)=\phi h(\rho)$ in our case is also
  found in the literature. The advantage of such a parametrization is
  that also the flow of general functions $\tilde{h}(\phi)$ not
  necessarily satisfying the symmetry constraints can be
  analyzed. This was first derived in \cite{Zanusso:2009bs} within the
  context of gravitational corrections to Yukawa systems. A detailed
investigation of the phase structure of the model in $d<4$ has been
performed in \cite{Vacca:2015nta}.

In the present work, it suffices to treat the gauge sector on the
perturbative level.  This approximation is justified because we are
only interested in the properties of the flow equations far above the
QCD scale. Near the electroweak scale threshold effects set in which
imply that all scalar self-interactions and all Yukawa couplings, as
well as the anomalous dimensions of the scalar and fermion field
freeze out before $g$ grows to large values at the typical IR energy
scale of QCD.

The perturbative one-loop beta function for the gauge coupling is given by
\begin{align}
 \pt\, g^2 = g^2 \eta_{\rmG}, \qquad \eta_{\rmG} = -\frac{g^2}{8\pi^2} \Big( \frac{11}{3}N_{\mathrm{c}} - \frac{2}{3}\Nf  \Big),
 \label{eq:flowgaugecoupling}
\end{align}
where $\Nf$ is the number of different quark flavors in the model. In order to describe a standard-model-like behavior, we choose $\Nf = 6$. Though our truncation contains  the top quark explicitly, we may add additional $SU(\Nc)$-invariant kinetic terms for the five other standard model quark flavors, but set their Yukawa couplings to zero. Thus, we consider the correct running of the pure strong sector at one-loop order while the flow equations for the Higgs-top sector are not affected by this modification. Due to the smallness of the Yukawa couplings of the other fermions, this is an acceptable approximation for the pure Higgs-top-QCD sector of the standard model, as has quantitatively been discussed in \cite{Eichhorn:2015kea}.

\subsection{(Extended) Mean-field approximation}
\label{sec:EMF}
The system of nonperturbative flow equations for the scalar \eqref{eq:flowPot} and the Yukawa potential \eqref{eq:flowYukawa} is a set of coupled nonlinear partial differential equations and thus nontrivial to solve. For analytical insights, mean-field as well as extended mean-field approximations have proven useful \cite{Gies:2013fua,Gies:2014xha}. Moreover, they allow for a good qualitative as well as quantitative description of the lower Higgs mass bound and of the stability issue of the scalar potential beyond polynomial approximations \cite{Borchardt:2016xju}.

For the mean-field approximation, we take only pure fermionic fluctuations into account and keep the wave function renormalizations fixed, $Z_i \to 1$ where the index $i$ labels the different fields. This approximation corresponds to a fermionic one-loop approximation and is directly related to the fermion determinant. In the language of flow equations, we neglect the anomalous dimensions, taking only terms with fermion lines in the loops into account, and furthermore replace the RG-improved full propagator at an RG scale $k$ by the propagator of the UV action $\Gamma_k^{(2)} \to \Gamma_\Lambda^{(2)}$ on the right-hand side of the flow equation. Thus, the flow equations for the dimensionful potentials read:
\begin{align*}
 \pt U_k(\rho) &= - 2 v_{d} d_{\gamma}\Nc \, k^d \, l_{0}^{(\rmF)d}\Big([2\rho H_\Lambda^{2}]/k^2 ; 0\Big), \\
 \pt H_k(\rho) &= 0,
\end{align*}
where $\rho=\phi^2/2$ is the dimensionful $\mathbb{Z}_2$ invariant. Both flow equations decouple in this approximation and can be integrated straightforwardly. For a convenient choice of the regulator (Litim regulator), the result is
\begin{align}
 \HMF_k(\rho) &= \HL(\rho), \\
 \UMF_k(\rho) &= \UL - \frac{d_{\gamma} \Nc (\Lambda^2-k^2)}{32 \pi^2}\rho\HL^2 \notag \\
              &\quad + \frac{d_{\gamma} \Nc}{16 \pi^2} \rho^2 \HL^4 \ln{\frac{\Lambda^2+2\rho\HL^2}{k^2+2\rho\HL^2}}.
\end{align}
If we introduced $N$ copies of the top quark, this approximation would become exact in the large-$N$ limit.

By contrast, the extended mean-field (EMF) approximation takes $1/N$ corrections into account and thus includes bosonic fluctuations on the same Gau\ss ian level. The scalar potential and the Yukawa potential read in the EMF approximation:
\small
\begingroup
{\allowdisplaybreaks
\begin{align}
 \UEMF_k(\rho) &= \UL + \frac{1}{64\pi^2} \Big[ M_\phi^2 (\Lambda^2-k^2) - M_{\phi}^4 \ln{\frac{\Lambda^2+M_\phi^2}{k^2+M_\phi^2}} \Big] \notag \\
              &\quad - \frac{d_\gamma\Nc}{64\pi^2} \Big[ M_\rmT^2 (\Lambda^2-k^2) - M_{\rmT}^4 \ln{\frac{\Lambda^2+M_\rmT^2}{k^2+M_\rmT^2}} \Big],  \\
 \HEMF_k(\rho) &= \HL(\rho) 
                  + \frac{1}{64\pi^2} \frac{\HL(\rho) \big(\HL(\rho) + 2\rho \HL'(\rho)\big)^2}{M_\phi^2-M_\rmT^2} \times \notag \\
               &\quad \times \Big[ \frac{(\Lambda^2-k^2)M_\phi^4}{(\Lambda^2+M_\phi^2)(k^2+M_\phi^2)} - 2M_\phi^2 \ln{\frac{\Lambda^2 + M_\phi^2}{k^2 + M_\phi^2}}  \notag \\
               &\qquad -  \frac{(\Lambda^2-k^2)M_\rmT^4}{(\Lambda^2+M_\rmT^2)(k^2+M_\rmT^2)} +  2M_\rmT^2 \ln{\frac{\Lambda^2 + M_\rmT^2}{k^2 + M_\rmT^2}} \Big] \notag \\
               &\quad + \frac{1}{64\pi^2} \big(3\HL'(\rho)+2\rho\HL''(\rho)\big) \times \notag \\
               &\quad \times \Big[ (\Lambda^2-k^2) + \frac{(\Lambda^2-k^2)M_\phi^4}{(\Lambda^2+M_\phi^2)(k^2+M_\phi^2)} \notag \\
               &\qquad - 2M_\phi^2 \ln{\frac{\Lambda^2 + M_\phi^2}{k^2 + M_\phi^2}} \Big] \notag \\
               &\quad - \frac{3+\xi}{24 \pi^2}g^2 \HL(\rho) \Big[ \frac{(\Lambda^2-k^2)M_\rmT^2}{(\Lambda^2+M_\rmT^2)(k^2+M_\rmT^2)} \notag \\
               &\qquad - 2\ln{\frac{\Lambda^2 + M_\rmT^2}{k^2 + M_\rmT^2}} \Big]. \label{eq:HEMF}
\end{align}
}%
\endgroup
\normalsize
Here, we have introduced the abbreviations $M_\phi^2 = \UL'(\rho) + 2\rho \UL''(\rho)$ and $M_\rmT^2 = 2 \rho \HL(\rho)^2$, which are field dependend mass terms for the scalar field and the fermion, respectively.

\section{Projecting onto the Yukawa functional}
\label{sec:projection}

In this section, we discuss a particularity in the extraction of the
nonperturbative flow equation of the Yukawa potential from the
Wetterich equation. In fact, different projections have been used in
the literature, but a comprehensive analysis and mutual comparison is
missing so far. This section necessarily is rather technical for
readers who are not familiar with the functional RG and might be
skipped. Once we have settled the issue in this section, it turns out
to play only a minor role for the phenomenological implications of
higher-order operators on the lower Higgs mass bound discussed in the
next section.

In order to project onto the Yukawa potential $h(\trho)$, we first project onto the derivative-free fermion bilinear. To be more precise, we take functional derivatives of the flow equation with respect to $\psib$ and $\psi$, then consider vanishing external momenta for these two fields, homogeneous field configurations for the scalar field, as well as vanishing other fields. 
Applying this projection rule on both sides of the Wetterich equation yields a flow equation for the operator $\hat\phi\, h(\hat\rho)$, which reads in dimensionless renormalized quantities
\begin{align}
 \pt [ \tilde\phi \, h(\trho) ] = \tilde\phi \, \beta_h(\trho).
\end{align}
The right-hand side of the Wetterich equation in this model is a function of the homogeneous scalar field which can be expressed in terms of the field $\tilde\phi$ times a function of the field invariant $\trho$ due to the symmetry of the model. This beta functional $\beta_h(\trho)$ is given on the right-hand side of Eq.~\eqref{eq:flowYukawa} for the present truncation \eqref{eq:EffectiveAverageAction}.

In order to investigate the running of the pure Yukawa function $h$, we are able to continue in two different ways. The obvious projection onto $h$ is given by factoring out the scalar field and identifing, see, e.g., \cite{Jungnickel:1995fp,Zanusso:2009bs,Pawlowski:2014zaa,Vacca:2015nta,Jakovac:2015kka},
\begin{align}
 \pt\, h(\trho) = \beta_{h}(\trho).
 \label{eq:projectionPawlowski}
\end{align}
On the other hand, we can project onto the 3-point function $\Gamma_{\phi\psib\psi}^{(3)}$ by taking an additional derivative with respect to the scalar field $\phi$, as one would naively do for a model with a field-independent Yukawa coupling \cite{Gies:2009hq,Gies:2009sv,Gies:2013pma,Gies:2013fua}. Thus, we finally obtain
\begin{align}
 \pt\, h(\trho) = \beta_{h}(\trho) + 2\trho \beta_{h}'(\trho) - 2\trho\, \pt h'(\trho).
 \label{eq:projectionGies}
\end{align}
Obviously Eq.~\eqref{eq:projectionPawlowski} and Eq.~\eqref{eq:projectionGies} coincide as long as no further approximation on the function $h(\trho)$ is maintained, as $\pt h_k=\beta_{h}$ solves Eq.~\eqref{eq:projectionGies} self consistently. This is expected because any projection onto the unique operator $\phi\psib\psi$ should result in a unique flow equation for the Yukawa potential $h$.

However, it is nontrivial to solve this nonlinear partial differential equation coupled to other differential equations, such as the flow equation for the scalar potential or the gauge coupling, and the algebraic equations of the anomalous dimensions. A standard approximation is to expand the potentials locally in a power series and to investigate the running of the various coefficients, reducing the system to a set of coupled ordinary differential equations. Such approximations have proven useful \cite{Berges:2000ew,Braun:2011pp,Braun:2012zq,Janssen:2014gea,Pawlowski:2014zaa,Braun:2014ata,Krippa:2014kra,Vacca:2015nta,Jakovac:2015kka}
especially if one is interested in local properties of the potentials such as the Higgs or the top mass being excitations on top of the electroweak minimum. Of course, global properties can usually not be resolved by these local approximations in field space and investigations of the full system seem to be indispensable to address questions of stability issues of the potentials, cf. for example \cite{Borchardt:2016xju}.

As soon as we use a finite-order approximation for the Yukawa function, differences in the flow equations \eqref{eq:projectionPawlowski} and \eqref{eq:projectionGies} are possible, depending on the approximations used. 
This can be illustrated by the following simple example. Suppose we expand the Yukawa potential in a power series,
\begin{align}
 h(\trho) = \sum_{n=0}^{N_h} \frac{y_{n}}{n!}(\trho - \trho_{0})^n,
 \label{eq:htrunc}
\end{align}
where also the expansion point $\trho_0$ can be chosen RG-scale dependent. A convenient choice is the actual scale-dependent minimum of the scalar potential $\kappa$. This can be either at vanishing field amplitude for the symmetric case or at some nontrivial value in case the system is in the SSB regime, defined by $u'(\kappa)=0$.
Choosing $\trho_0 = \kappa$, the coefficients of the expansion have a natural interpretation as couplings between the scalar field and the fermion. The absolute value $y_{0}$ corresponds to the usual field-independent Yukawa coupling $y_0\, \phi\psib\psi$ and the other couplings correspond to higher-dimensional generalized Yukawa couplings.

The running of the various coefficients of the expanded Yukawa function can be extracted straightforwardly from the beta functional $\beta_{h}$ by appropriate derivatives with respect to the scalar field. 
In general, the RG evolution of these parameters are given by either
\begin{align}
 \pt y_n &= \pt h^{(n)}|_{\rho=\kappa} + h^{(n+1)}(\kappa) \pt \kappa \notag \\
 &= \beta_h^{(n)}(\kappa) + y_{n+1} \pt\kappa
 \label{eq:projectionPawlowski2}
\end{align}
or
\begin{align}
\pt y_n &= \beta_h^{(n)}(\kappa) + \frac{2\kappa}{1+2n} (\beta_h^{(n+1)}(\kappa) - \pt h^{(n+1)} |_\kappa) \notag \\
&\quad + y_{n+1} \pt \kappa
\label{eq:projectionGies2}
\end{align}
depending on the projection rule given by Eq.~\eqref{eq:projectionPawlowski} or Eq.~\eqref{eq:projectionGies}, respectively. The superscript $^{(n)}$ indicates the $n$-th derivative with respect to $\tilde\rho$.
It is obvious that both flow equations coincide in the SYM regime, $\kappa = 0$. 
In the SSB regime both flow equations agree with each other as the difference in both projection rules, $2\kappa (\beta_h^{(n+1)}(\kappa) - \pt h^{(n+1)}_k|_\kappa)$, vanishes identically order by order for any coupling $y_{n}$ as long as the series of the Yukawa function $h$ is not truncated.

Nevertheless, as soon as any finite-order approximation for the Yukawa function is considered, the running of any $y_n$ does not capture the full back-reaction of the entire generalized Yukawa potential.  
For a moment, let us consider the simplest possible truncation of the given local approximation in field space \eqref{eq:htrunc}, $h(\tilde\rho)=y_0$. 
Thus the right-hand side of the flow equation becomes only a function of $y_0$ (as well as scalar self couplings, which we will suppress in the following), $\beta_{y_0}(y_0) \equiv \pt y_0 = \beta_h(\kappa)$. 
By contrast, the RG evolution of the Yukawa coupling $y_0$ will get corrections from higher-dimensional operators beyond this simple truncation as the beta functional $\beta_{h}$ depends on the first and second derivative of $h$ with respect to $\tilde\rho$. Thus, the flow equation of $y_{0}$ will explicitly depend on $y_{1}$ and $y_{2}$ as soon as the truncation order $\Nh$ is increased, $\beta_{y_0}(y_0,y_1,y_2)$. 
In general, the running of any Yukawa type coupling will be a function $\beta_{y_n}(y_0,\cdots,y_n,y_{n+1},y_{n+2})$, leading to an infinite tower of nonlinear coupled ordinary differential equations. Truncating this tower leads to artifacts which have to be studied carefully in any functional RG investigation.

As already mentioned any finite-order approximation in addition leads to different runnings of the Yukawa couplings in the SSB regime given by either Eq.~\eqref{eq:projectionPawlowski} or Eq.~\eqref{eq:projectionGies}.
This can be exemplified by the running of the absolute term $y_0$ for simplicity. 
Thus, for the current truncation ($\Nh=0$) the flow equation for the Yukawa coupling $y_0$ within the different approximations reads
\begin{align}
 \pt y_0 &= \beta_h(\kappa), \quad \text{or} \label{eq:projectionPawlowski3} \\ 
 \pt y_0 &= \beta_h(\kappa) + 2\kappa\, \beta_h'(\kappa).   \label{eq:projectionGies3}
\end{align}

\begin{figure}[t!]
\centering
\begin{minipage}{0.05\textwidth}
 $\pt y_0: $
\end{minipage}
\hfill
\begin{minipage}{0.12\textwidth}
\includegraphics[width=1\textwidth]{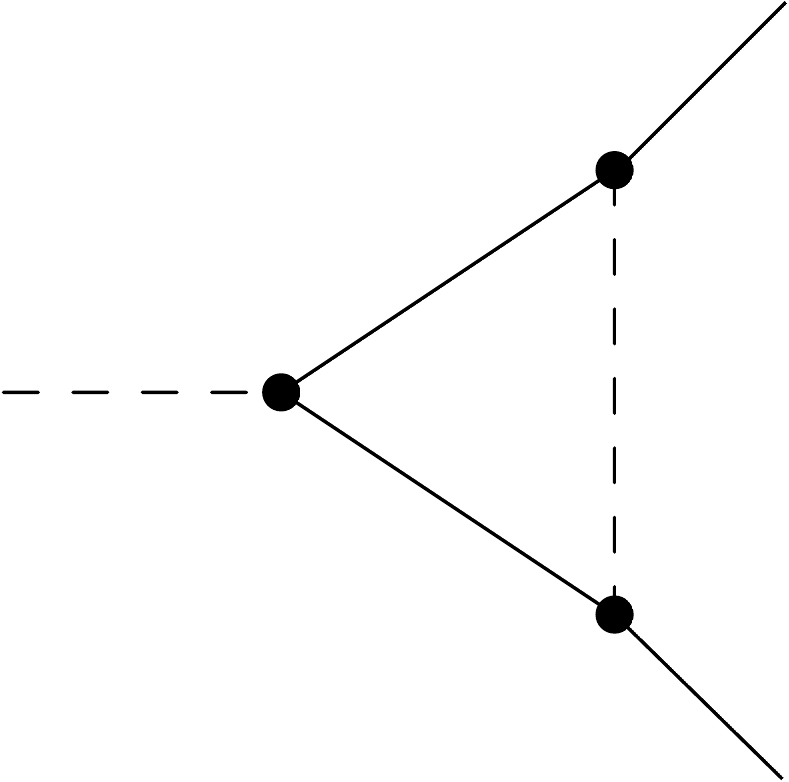}
\end{minipage}
\hfill
\begin{minipage}{0.12\textwidth}
\includegraphics[width=1\textwidth]{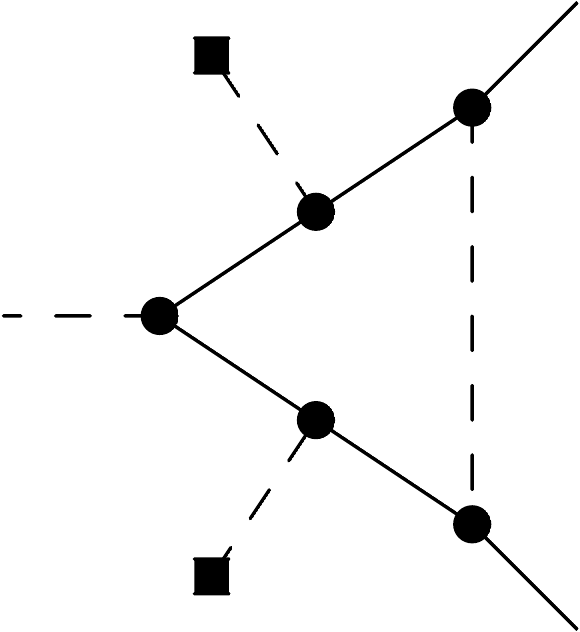}
\end{minipage}
\hfill
\begin{minipage}{0.12\textwidth}
\includegraphics[width=1\textwidth]{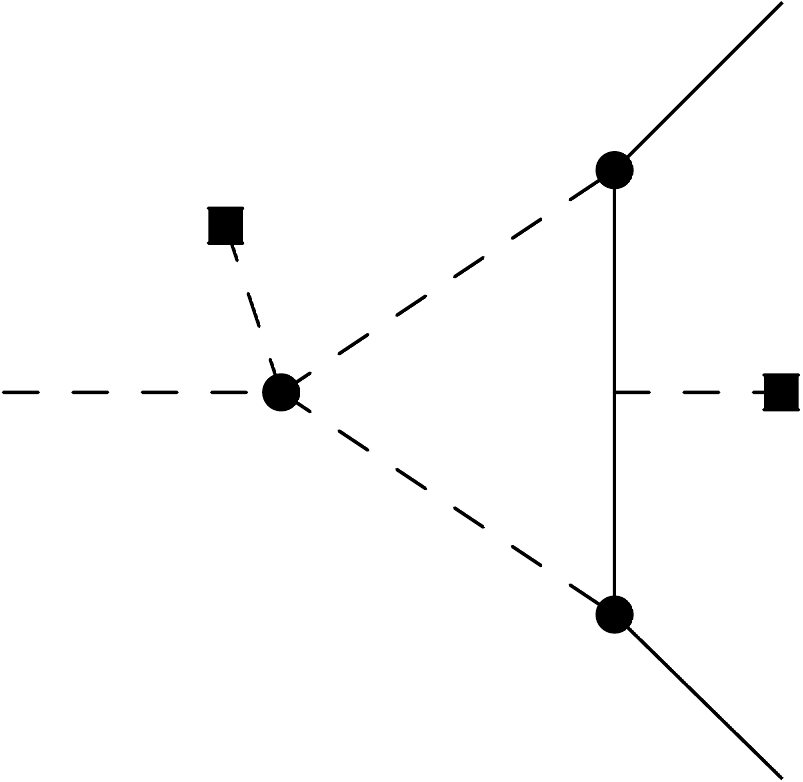}
\end{minipage} 
\caption{Diagrammatic contributions to the running of the Yukawa coupling $y_{0}$. Filled squares denote couplings to the condensate $\sim \sqrt{2\kappa}$, i.e., insertions of the expectation value of the scalar field. These contribute to the flow in the SSB regime if the second projection rule Eq.~\eqref{eq:projectionGies} is used.}
\label{fig:Feyn1}
\end{figure}

\begin{figure}[t!]
\centering
\begin{minipage}{0.05\textwidth}
(a)
\end{minipage}
\begin{minipage}{0.11\textwidth}
\includegraphics[width=1\textwidth]{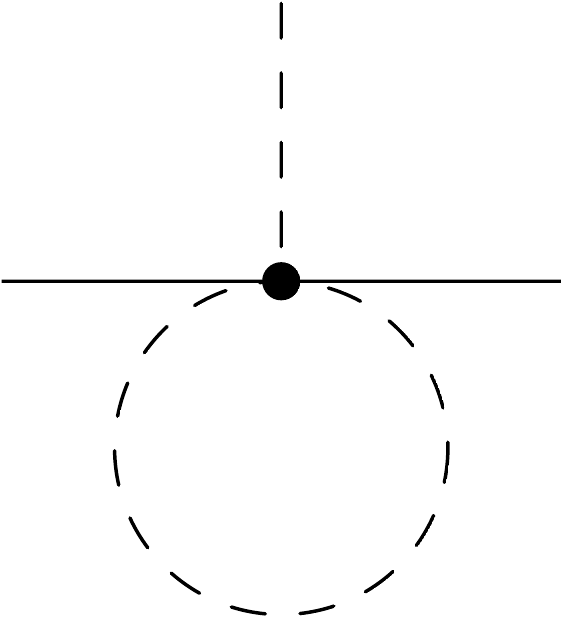}
\end{minipage}
\\[0.02\textwidth]
\begin{minipage}{0.05\textwidth}
 (b)
\end{minipage}
\begin{minipage}{0.12\textwidth}
\includegraphics[width=1\textwidth]{Figures/triangle2c.pdf}
\end{minipage}
\begin{minipage}{0.05\textwidth} 
$\to$ 
\end{minipage}
\begin{minipage}{0.1\textwidth}
\includegraphics[width=1\textwidth]{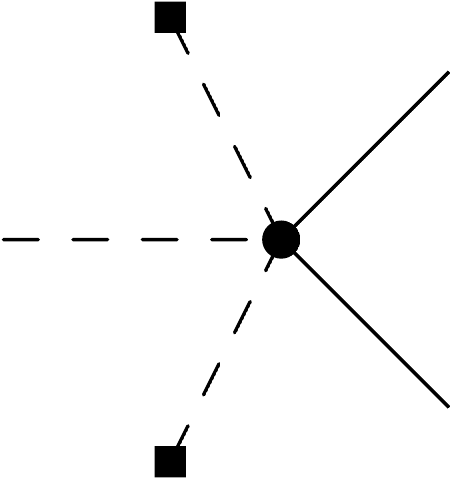}
\end{minipage} 
\caption{(a) Example for an additional contribution, which is proportional to $y_1$, to the flow of the Yukawa coupling $y_0$, if higher-order operators of the form $\rho^n \phi\psib\psi$ are taken into account. 
(b) Parts of these higher-order couplings were already considered in the flow equation of $y_{0}$ in Eq.~\eqref{eq:projectionGies2}. The triangle diagrams with external expectation-value insertions (filled squares) in Fig.~\ref{fig:Feyn1} are now included in the running of the three-scalar-two-fermion vertex by restricting two of the scalar legs to couple to the expectation value. A similar translation holds for the third triangle diagram which has two scalar lines in the loop in Fig.~\ref{fig:Feyn1}.}
\label{fig3:Feyn2}
\end{figure}

In order to illustrate this difference between both projections in the SSB regime, i.e., the $2\kappa \beta_h'(\kappa)$ term, we give a diagrammatic interpretation of these flow equations in Fig.~\ref{fig:Feyn1}. Whereas only the first diagram contributes to the flow of the Yukawa coupling $y_0$ corresponding to Eq.~\eqref{eq:projectionPawlowski3}, the second and third diagram parametrize the additional contributions in Eq.~\eqref{eq:projectionGies3}. These additional contributions correspond to expectation-value insertions of the scalar field and effectively encode parts of the neglected higher-order terms in the expansion \eqref{eq:htrunc} as can be seen by the external legs with filled squares. To be more precise, the second and third diagram in Fig.~\ref{fig:Feyn1} provide corrections to the running of $y_{0}$ from an effective three-scalar-two-fermion vertex where two of the external scalar legs are attached to interactions with the condensate $\sqrt{2 \kappa}$. In general, these contributions would naturally be included in the beta functional $\beta_h(\kappa)$, and thus in the running of $y_0$, if we included the $y_1$ coupling in our truncation, see Fig.~\ref{fig3:Feyn2}.

Increasing the truncation order to $N_{h}=1$ leads to a full inclusion of the $y_1$ contribution to the flow equation of $y_{0}$ for both projection strategies, $\pt y_{0} = \beta_{h}(\kappa)=\beta_{y_{0}}(y_{0},y_{1})$. 
The former additional and now redundant contribution $2\kappa\, \beta_{h}(\kappa)$ in Eq.~\eqref{eq:projectionGies3} will be exactly canceled due to the nonvanishing $\pt h'|_{\kappa} = \beta_{h}'(\kappa) + \frac{2\kappa}{3}(\beta_{h}''(\kappa)-\pt h''|_{\kappa})$ beyond $\Nh=0$. Inserting this relation into Eq.~\eqref{eq:projectionGies2} for $n=0$ leads to a flow equation for $y_{0}$ in which a new contribution arise, given by $\frac{4}{3}\kappa^2  \beta_{h}''(\kappa)$ in the current truncation. 
This time, the additional term encodes parts of the negelcted contribution from the next higher coupling in the expansion $y_{2}$.
Note, that also the flow equation of $y_{1}$ consistently gets a modification from the five-scalar-two-fermion vertex which reads $\frac{2\kappa}{3}\beta_{h}''(\kappa)$.

The same consideration holds for any higher-order approximation of the Yukawa function. Both projection rules  
include the full dependence on the various generalized Yukawa couplings up to $y_{\Nh}$ for every $y_{n\leq\Nh}$ encoded in $\beta_{h}^{(n)}(\kappa)$. However, the projection prescription \eqref{eq:projectionGies} takes effectively parts of the neglected $y_{\Nh+1}$ coupling into account. This can be seen as Eq.~\eqref{eq:projectionGies2} can be written in a form in which the additional term in brackets is explicitly resolved
\begin{align}
 \pt y_{n}  &=  \beta_{h}^{(n)}(\kappa)  -  \Bigg( \prod_{l=0}^{\Nh-n} \frac{-2\kappa}{1+2(n+l)} \Bigg)  \beta_{h}^{(\Nh+1)}(\kappa) \notag \\  
 &\quad +  y_{n+1}\pt \kappa.
 \label{eq:projectionGies4}
\end{align}
%
Whether this leads to a quantitatively more accurate approximation
depends highly on the given problem and cannot be answered a
priori. For instance note that $y_{0}$ gets a correction from
couplings $y_{3 \leq n\leq \Nh}$ due to the second term on the
right-hand side even if $\beta_{y_{0}}$ should explicitly depend only
on $y_{0}$, $y_{1}$, and $y_{2}$. All other contributions from higher-dimensional operators should effectively be incorporated in $y_{1}$
and $y_{2}$. As we also work with polynomial epxansions in the
following, we monitor the differences between the projections.

\subsection{Extended mean-field analysis}
After these general considerations on the different projection schemes for the Yukawa couplings, let us investigate their consequences on the lower Higgs mass bounds in the present Yukawa models. Therefore, we begin with the extended mean-field calculation and afterwards turn to the full flow equations. In order to keep the technical part of the calculation transparent, we first restrict the discussion only to the gauge sector induced running of the Yukawa function in this subsection. This can already be justified from a phenomenological point of view by the fact that the gauge boson contributions dominate the running of the Yukawa sector in the standard model. To be more precise, we consider only the last line ${\sim}g^2$ on the right-hand side of the flow equation \eqref{eq:flowYukawa} for the Yukawa coupling. 

As shown in Sec. \ref{sec:EMF}, we are able to integrate the beta functional for a suitable regulator function by analytical means, see Eq.~\eqref{eq:HEMF}. The effective Yukawa potential can be obtained in the limit $k\to 0$, corresponding to integrating out all quantum fluctuations. Thus, the gauge boson induced effective potential reads,
\begin{align}
 \HEMF_{\mathrm{eff}}(\rho) &= \HEMF_{k=0}(\rho) \notag \\
 &= H_{\Lambda}(\rho)  -  \frac{3+\xi}{24\pi^2} g^{2} H_{\Lambda}(\rho) \times \notag \\
 &\qquad \times \bigg[ \frac{\Lambda^{2}}{\Lambda^{2} + M_\rmT^{2}}  -  2\ln \left(1+\frac{\Lambda^2}{M_\rmT^2} \right)\bigg].
\end{align}
We would like to emphasize again that this is the unique result for the full $\rho$-dependent (gauge-induced) integrated EMF effective Yukawa function. Since no approximation in field space has been performed, the result is independent of the projection scheme as the additional contributions in Eq.~\eqref{eq:projectionGies} cancel each other.

Expanding the effective Yukawa potential in a Taylor series around the vacuum expectation value of the effective scalar potential, $\rho_{0}=\lim_{k\to 0}k^{2}\kappa$, and extracting the usual effective Yukawa coupling, i.e., the absolute term $y_{0,\mathrm{eff}}$, we obtain according to the first projection scheme,
\begin{align}
 y_{0,\mathrm{eff}} &= \HEMF_{\mathrm{eff}}(\rho_{0}) \notag \\ 
 &= y_{0,\Lambda}  -  \frac{3+\xi}{24\pi^2} g^{2} y_{0,\Lambda} \bigg[ 1 - 2\ln \frac{\Lambda^{2}}{2\rho_{0}\, y_{0,\Lambda}^{2}} \bigg] + \mathcal{O}\bigg(\frac{\rho_{0}}{\Lambda^{2}}\bigg).
 \label{eq:y0eff1}
\end{align}
Here, we have assumed that the cutoff $\Lambda$ of the effective theory is much larger than the other scales in the theory set by the electroweak vacuum, $\Lambda^{2}\gg \rho_{0}$. 
Note that $y_{0,\eff}$ does not depend on $y_{1,\eff}$ or any higher-order coefficient within this approximation as they are either suppressed by a suitable power of $\Lambda$ or neglected due to the fact that we only investigate the gauge-induced part. 
Thus, we obtain a rather fast convergence with respect to $\Nh$ for the gauge-induced effective Yukawa coupling within the EMF approach for this projection scheme. $\Nh=0$ captures already all the relevant contributions. 
Similar conclusions hold also for the neglected radiative corrections coming from the fermion-scalar interactions in Eq.~\eqref{eq:flowYukawa}. Taking these also into account results amongst others in an additional $\mathcal{O}(1)$ term ${\sim}y_{1,\Lambda}$ to $y_{0,\eff}$ coming from the contribution $H_{\Lambda}'(\rho)$ in Eq.~\eqref{eq:flowYukawa}. Nonetheless, all $y_{n\geq 2}$ contributions are suppressed for any value of $\Nh$.

However, using Eq.~\eqref{eq:projectionGies} and the finite order expansion \eqref{eq:htrunc}, we get a correction term according to Eq.~\eqref{eq:projectionGies4}, encoding some of the truncated parts of the series. Within our easy-to-calculate approximation we can perform the necessary steps analytically to any order $\Nh$ in the truncation. The effective Yukawa coupling reads,
\begin{align}
 y_{0,\mathrm{eff}} 
 &= y_{0,\Lambda}  -  \frac{3+\xi}{24\pi^2} g^{2} y_{0,\Lambda} \bigg[ 1 - 2\ln \frac{\Lambda^{2}}{2\rho_{0}\, y_{0,\Lambda}^{2}}  \notag \\ 
 &\quad +  \frac{4(2\Nh)!!}{(2\Nh+1)!!} \bigg]  + \mathcal{O}\bigg(\frac{\rho_{0}}{\Lambda^{2}}\bigg),
 \label{eq:y0eff2}
\end{align}
where the last term in the square bracket mimics parts of the higher-order corrections.
First of all, we notice that this expression converges to Eq.~\eqref{eq:y0eff1} for $\Nh \to \infty$ as it should, because in that limit both projection rules have to agree. However, the convergence is rather slow as it behaves as $\Nh^{-1/2}$ for large $\Nh$. Analogous to Eq.~\eqref{eq:y0eff1}, Eq.~\eqref{eq:y0eff2} does not explicitly depend on $y_{n\geq 1}$ if we neglect terms which are suppressed by at least $\Lambda^{2}$. Thus, we get a rather slow convergence of our results by increasing $\Nh$ in contrast to Eq.~\eqref{eq:y0eff1}, as the additional contributions encoding the higher-order operators will overshoot their actual impact on the gauge-induced running of $y_{0}$.

Fixing a physical parameter such as the top mass is technically implemented by a renormalization condition. We impose this renormalization condition for the effective Yukawa coupling $y_{0,\eff}=\HEMF_{\eff}(\rho_{0})$, given by $\mtop^{2} = 2\rho_{0} y_{0,\eff}^{2}$. 
Thus, the different projection schemes only lead to a redefinition of the bare Yukawa coupling $y_{0,\Lambda}$ which seems insignificant within the pure Yukawa sector at first sight. We solve either Eq.~\eqref{eq:y0eff1} or Eq.~\eqref{eq:y0eff2} with respect to $y_{0,\Lambda}$ for $\mtop=173$ GeV and $\rho_{0}=246^{2}/2$ GeV$^{2}$ using the Landau gauge $\xi=0$. For a small cutoff scale $\Lambda=1$ TeV as well as for small $\Nh$ and for gauge couplings of $\mathcal{O}(1)$, we observe  a deviation of a few percent between the two schemes. This difference decreases for larger cutoffs or rather larger $\Nh$, of course.

However, this shift of $y_{0,\Lambda}$ and thus of the underlying microscopic theory will have an explicit impact on the scalar sector. For instance, the lower Higgs mass bound is mainly dominated by top fluctuations, such that even small deviations of the bare Yukawa coupling can have a significant influence on the bound. The lower Higgs mass bound within quartic bare Higgs potentials is approached for vanishing quartic couplings, $U_{\Lambda}(\rho) = \mL \rho$. For this case, the mean-field approximation for the scalar potential $\UMF$ coincides with the extended mean-field approximation $\UEMF$ up to an irrelevant constant shift. Therefore, an analytic expression for the lower Higgs mass bound can be worked out \cite{Gies:2013fua},
\begin{align}
 \mH^{2} &=  2\rho_{0} {\UMF_{\eff}}''(\rho_{0})  \notag \\ 
 &= \frac{\rho_{0}y_{0,\Lambda}^{4}}{2\pi^{2}} \left[ 2 \ln \frac{\Lambda^{2}}{2\rho_{0}y_{0,\Lambda}^{2}} -3 \right]  + \mathcal{O}\bigg(\frac{\rho_{0}}{\Lambda^{2}}\bigg).
\end{align}
As can be seen by this formula for the lower Higgs mass bound, an increase of $y_{0,\Lambda}$ by say 5\% leads up to 20\% larger Higgs masses. Thus any computation within the second projection scheme for the Yukawa coupling has to be treated with care. For large $\Nh$ the Higgs masses converge but there might be noticeable deviations from this value for small $\Nh$.

\subsection{Nonperturbative flow}
The extended mean-field analysis was valuable to get first insights in the convergence behavior of the truncation of the flow equations for the present Yukawa models. The next step is to investigate the impact of the two different projection schemes on the nonperturbative flow equations with all mutual back reactions between the bosonic and fermionic degrees of freedom. Of course, solving the full system of coupled partial differential equations would be the ideal case. Recently, numerical solvers based on pseudo-spectral methods have been developed to investigate the properties of global solutions of functional flow equations in various systems \cite{Borchardt:2015rxa,Borchardt:2016pif,Borchardt:2016kco,Knorr:2016sfs}. However, our primary interest is the lower Higgs mass bound resulting from the RG flow. For this, we only need local information in field space around the electroweak vacuum. Thus, it suffices to consider a Taylor expansion of the scalar potential about the flowing electroweak minimum
\begin{align}
 u(\trho) = \sum_{n=1}^{\Nu} \frac{\lambda_n}{n!} (\trho - \kappa)^{n}.
 \label{eq:utrunc}
\end{align}
In the symmetric regime ($\kappa=0$), we identify $\lambda_{1}$ with
the dimensionless running mass term. In the SSB regime, this parameter
vanishes as we expand the potential around the nonvanishing
scale-dependent minimum $\kappa$. The quartic Higgs self-coupling is
given by $\lambda_{2}$. The other couplings parametrizing the
potential correspond to higher-dimensional scalar self-interactions.
Similarly, we expand the Yukawa potential $h$ in a power series
according to Eq.~\eqref{eq:htrunc} with $\trho_{0}=\kappa$.

As the lower Higgs mass bound is associated with small scalar self-couplings, the scalar sector is mostly dominated by top fluctuations in ungauged as well as gauged Yukawa theories. Fermionic fluctuations drive the system toward the SSB regime and can induce chiral symmetry breaking. Thus, a typical flow for a weakly interacting bare scalar sector has to start in the symmetric regime with a suitable bare scalar mass term $\lambda_{1,\Lambda}$ in order to achieve a desired scale separation between the cutoff scale $\Lambda$ and the Fermi scale $v=\sqrt{2\rho_{0}} = 246$ GeV. The fermionic fluctuations drive the system into the SSB regime near the electroweak scale and a vacuum expectation value for the scalar field builds up. At this point, we switch from the symmetric ($\kappa=0$) to the SSB ($\kappa \neq 0$) parameterization of the scalar and the Yukawa potential and investigate the running of the respective couplings. The nonvanishing vacuum expectation value induces masses for the top and the Higgs excitation, leading to a decoupling of modes, and the flow for the potentials freezes out completely.

\begin{figure*}[t!]
\centering
\hfill
\includegraphics[width=0.4\textwidth]{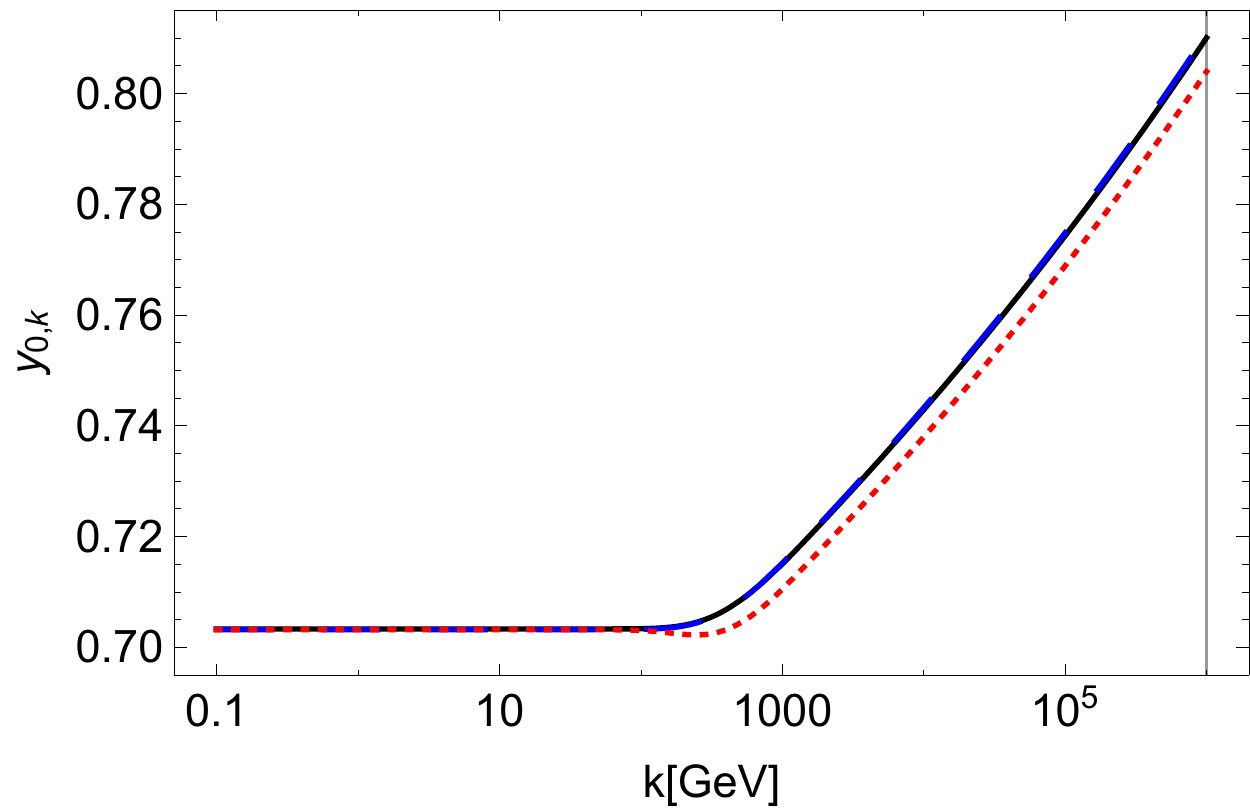}\hfill
\includegraphics[width=0.4\textwidth]{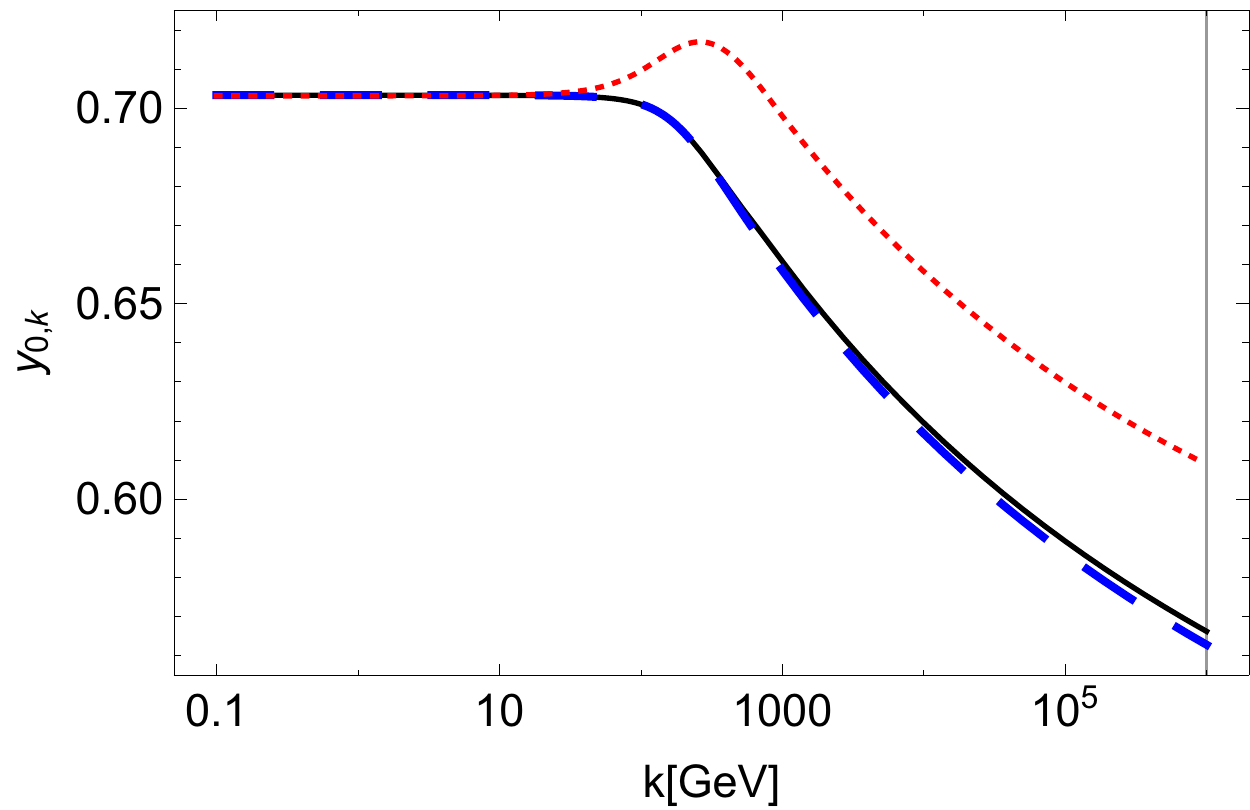}\hfill\quad
\caption{Example flows for the Yukawa coupling $y_{0}$ for different truncations. The left panel depicts the flow in the ungauged model while the right plot shows the RG running within the gauged Higgs-Yukawa toy model. The blue dashed line and the black solid curve correspond to a truncation $\Nh=0$ and $\Nh=1$ respectively and the flow equations are extracted according to Eq.~\eqref{eq:projectionPawlowski}. The red dotted curve shows the RG flow resulting from Eq.~\eqref{eq:projectionGies} for $\Nh=0$.}
\label{fig:Running-y0}
\end{figure*}

Due to the freeze-out of the scalar and Yukawa couplings, it is
convenient to choose a deep IR scale as renormalization point. Our
parametrization of the potentials allows to extract the
phenomenologically relevant information in a straightforward
manner. The renormalization conditions for the vacuum expectation
value, the top mass, and the Higgs mass read $U_{\eff}'(\rho_{0}) =
0$, $\mtop^{2} = 2 \rho_{0} h_{\eff}(\rho_{0})^{2}$, and $\mH^{2} =
2\rho_{0} U_{\eff}''(\rho_{0})$, respectively. Here $\rho_{0}$
denotes the renormalized field value at the minimum of the scalar
potential. Though we only consider toy models in the present work, we
choose $\mtop = 173$GeV and $v = 246$GeV to mimic standard model
properties.\footnote{In standard-model phenomenology, it is important
  to fix the physical parameters in an appropriate scheme that allows
  to make contact with experiment. For instance, the top mass can
  differ significantly for different schemes. As the regularization
  procedure of the functional RG comes with its own scheme, the
  parameter fixing is not straightforward. In \cite{Eichhorn:2015kea},
  it has been shown that a quantitatively acceptable fixing can be
  achieved by matching the functional RG flows to those of
  conventional schemes at the TeV scale. As the present toy models do
  not allow for a direct quantitative comparison with the
  standard model anyway, we use a straightforward fixing in terms of the
  parameters of the effective potential for simplicity. } The Higgs
mass is treated as a free parameter. As a consequence of the RG flow,
it is not an arbitrary parameter but depends on the maximal UV extent
$\Lambda$ of the effective field theory as well as the precise form of
the UV action, $\mH = \mH[S_{\Lambda},\Lambda]$.

Likewise, it is possible to map the renormalization conditions in the IR on the UV initial parameters of the theory with the aid of the flow equations. 
In practise, we fine-tune the bare scalar mass term to match the phenomenological requirement of an IR vacuum expectation value of $246$ GeV. This fine-tuning of an RG relevant operator corresponds to the hierarchy problem of separating the UV cutoff from the electroweak scale. Further, we vary the bare Yukawa coupling $y_{0,\Lambda}$ to obtain a top mass of $\mtop = 173$ GeV.

In the gauged model, we have to determine initial conditions also for the flow equation of the strong coupling constant. 
In order to solve the  flow equation $\pt g^{2}$, we choose an initial value at the UV cutoff  scale $\Lambda$ such that we obtain an IR value at the mass scale of the Z boson of $g^{2}(k = m_{\mathrm{Z}}) = 4 \pi \alpha_{\mathrm{s}}(k = m_{\mathrm{Z}}) = 4 \pi \cdot 0.118$. 
The simple perturbative flow of the strong coupling constant does not include threshold effects for the top quark as the other nonperturbative flow equations do. We parametrize this threshold effect by using $\Nf=5$ in Eq.~\eqref{eq:flowgaugecoupling} between the Z-boson and top-mass scale instead of $\Nf=6$ above the top-mass scale.

The convergence of the polynomial truncations can be tested by successively increasing the number of couplings $\Nu$ and $\Nh$ parameterizing the scalar and Yukawa potential, respectively. For $\Nu$ this was investigated in \cite{Gies:2013fua} where truncations larger than $\Nu=4$ do not lead to any significant change of the derived Higgs masses. The same outcome can be confirmed for any fixed order of $\Nh$. Thus, for the following test of convergence of the two projection schemes, we fix $\Nu=4$ and study the influence of different $\Nh$ on the resulting Higgs masses.

In Fig.~\ref{fig:Running-y0}, an example flow of $y_{0}$ for the gauged as well as the ungauged Yukawa model is plotted for different truncations for $\Lambda = 10^{6}$ GeV and $\lambda_{n\geq 2} = 0$. For the ungauged model, we consider the nonperturbative flow equations~\eqref{eq:flowPot}-\eqref{eq:anomalFermion} provided in Sec.~\ref{sec:flow} for $g=0$ and $\Nc = 1$. The blue dashed line depicts the flow for $\Nh = 0$ for the first projection scheme according to Eq.~\eqref{eq:projectionPawlowski}. The black solid line shows the flow including corrections from $y_{1}$ with initial condition $y_{1,\Lambda} = 0$. For all larger truncations $\Nh > 1$ ($y_{n\geq 1, \Lambda} = 0$), no deviation from this black curve is visible by eye. This implies a rather fast convergence of the resulting IR Higgs masses within the first projection scheme similar to the extended mean-field approach. In fact, the difference of the Higgs masses obtained for $\Nh=0$ and $\Nh=1$ for the first projection scheme is $0.8$ GeV ($<1\%$ effect) for the ungauged, and $4.0$ GeV ($\approx 3\%$ effect) for the gauged model. No deviation from the $\Nh=1$ result can be observed within our numerical precision for the ungauged model by increasing the truncation order $\Nh$. For the gauged model the difference is less than $0.3$ GeV for $\Nh=2$ and beyond no deviation is observed.

The flow of $y_{0}$ extracted using the second projection of Eq.~\eqref{eq:projectionGies} for $\Nh=0$ is plotted as red dotted line in Fig.~\ref{fig:Running-y0}. 
Indeed, the additional contribution $2\kappa\,\beta'(\kappa)$ mimics parts of the effect of the higher-order coupling $y_{1}$ in the SSB regime qualitatively. For the ungauged model the Yukawa coupling $y_{0}$ is slightly smaller during the flow compared to the previous $\Nh=0$ case of the other projection scheme, c.f. blue dashed line. The flow of the Yukawa coupling is essentially dominated by $y_{0}$ itself within this model. Thus, it decreases from the UV towards the IR as  $\pt y_{0} \sim y_{0}^{3}$ with some positive scale-dependent coefficient. The higher-dimensional coupling $y_{1}$ contributes with the opposite sign to the running of $y_{0}$. Therefore, it relaxes the decrease of $y_{0}$ only slightly from a UV to IR perspective as the RG flow generates a small but positive $y_{1}$ in the UV if $y_{1,\Lambda} = 0$.

Within the second projection scheme, this relaxation is realized by
terms with a sign opposite to that of the primary quantum fluctuations
corresponding to the first diagram in Fig.~\ref{fig:Feyn1} and
containing the threshold function $l_{1,1}^{(\rmF\rmB)}$. These terms
are generated as soon as the derivative with respect to the scalar
field invariant hits one of the arguments of the threshold function

\begin{align}
 &2\kappa\, \partial_{\trho} l_{1,1}^{(\rmF\rmB)d}(2\trho y_{0}^{2}, u'+2\trho u'')|_{\trho=\kappa} \notag \\
 &\quad=  - 4 \kappa y_{0}^{2}\, l_{2,1}^{(\rmF\rmB)d}(2\kappa y_{0}^{2}, 2\kappa \lambda_{2}) \notag \\ 
 &\qquad - (6\kappa \lambda_{2} + 4\kappa^{2}\lambda_{3}) \, l_{1,2}^{(\rmF\rmB)d}(2\kappa y_{0}^{2}, 2\kappa \lambda_{2}).
 \label{eq:additional}
\end{align}
Their contribution to the flow is slightly stronger than the inclusion of the $y_{1}$ coupling within the first projection scheme leading to a smaller bare Yukawa coupling $y_{0,\Lambda}$ as can be seen in the left panel of Fig.~\ref{fig:Running-y0} and thus to a smaller Higgs mass in the scalar sector. The deviation from the converged value within the first projection scheme is $0.7$ GeV such that the second projection scheme gives a reasonable estimate for the higher-order contributions.

However, the fact that both higher-order approximations, $\Nh=1$ for the first and $\Nh=0$ for the second projection scheme, lead to almost similar values for the Higgs mass in the ungauged model is a subtle effect. In principle, the running of $y_{0}$ is modified by the $y_{1}$ coupling during the entire RG flow. The effective encoding of parts of the higher-order contributions in the second projection scheme takes place only in the SSB regime. 
Within the SYM regime, both projection rules lead to the same flow equations. Thus, we only expect a small difference between both as the system is mainly in the SYM regime during the flow from the UV towards the Fermi scale.

\begin{figure*}[t!]
\centering
\includegraphics[width=0.32\textwidth]{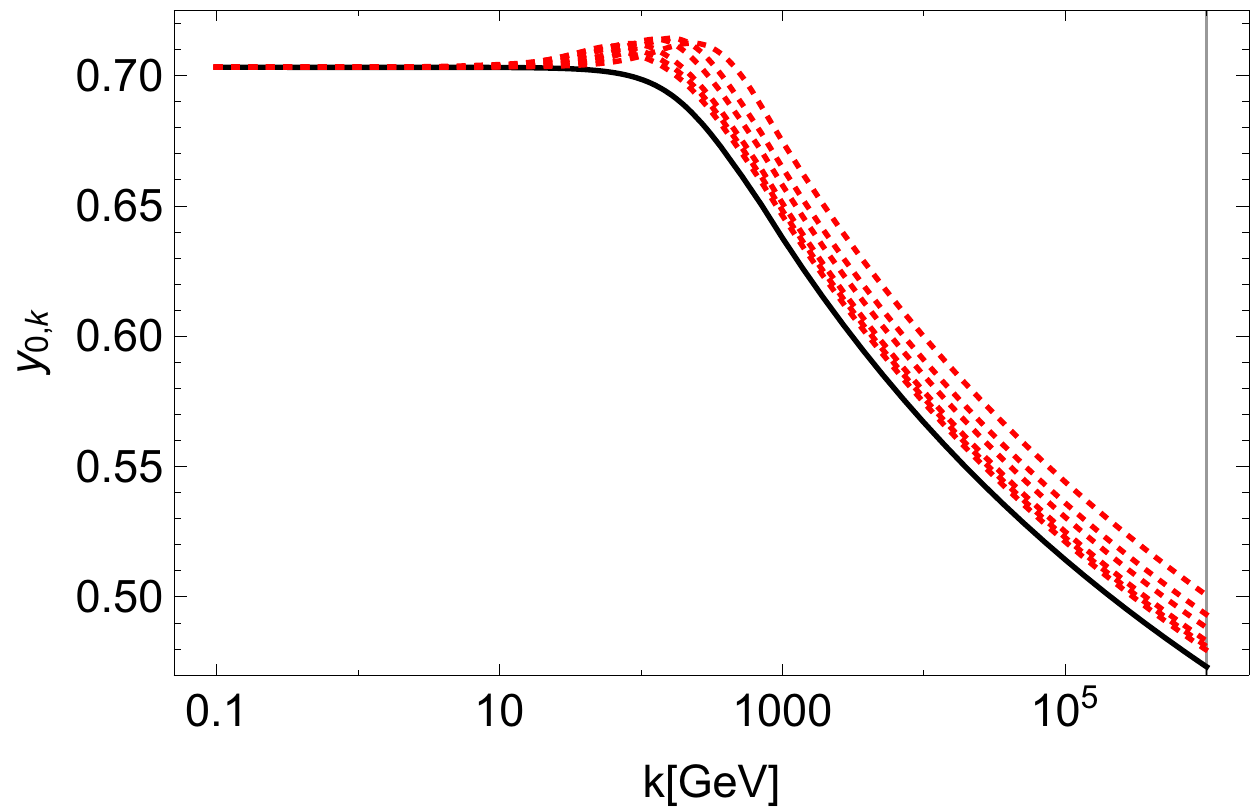}\hfill
\includegraphics[width=0.32\textwidth]{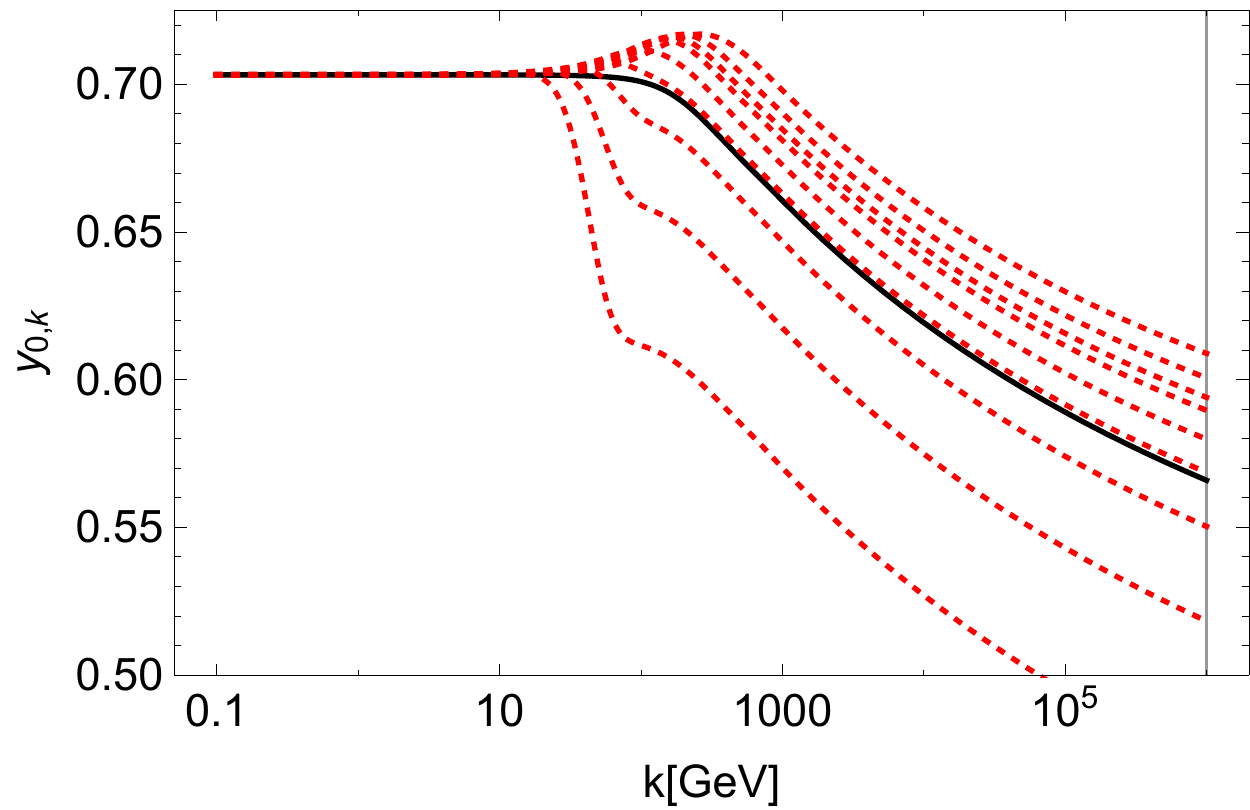}\hfill
\includegraphics[width=0.32\textwidth]{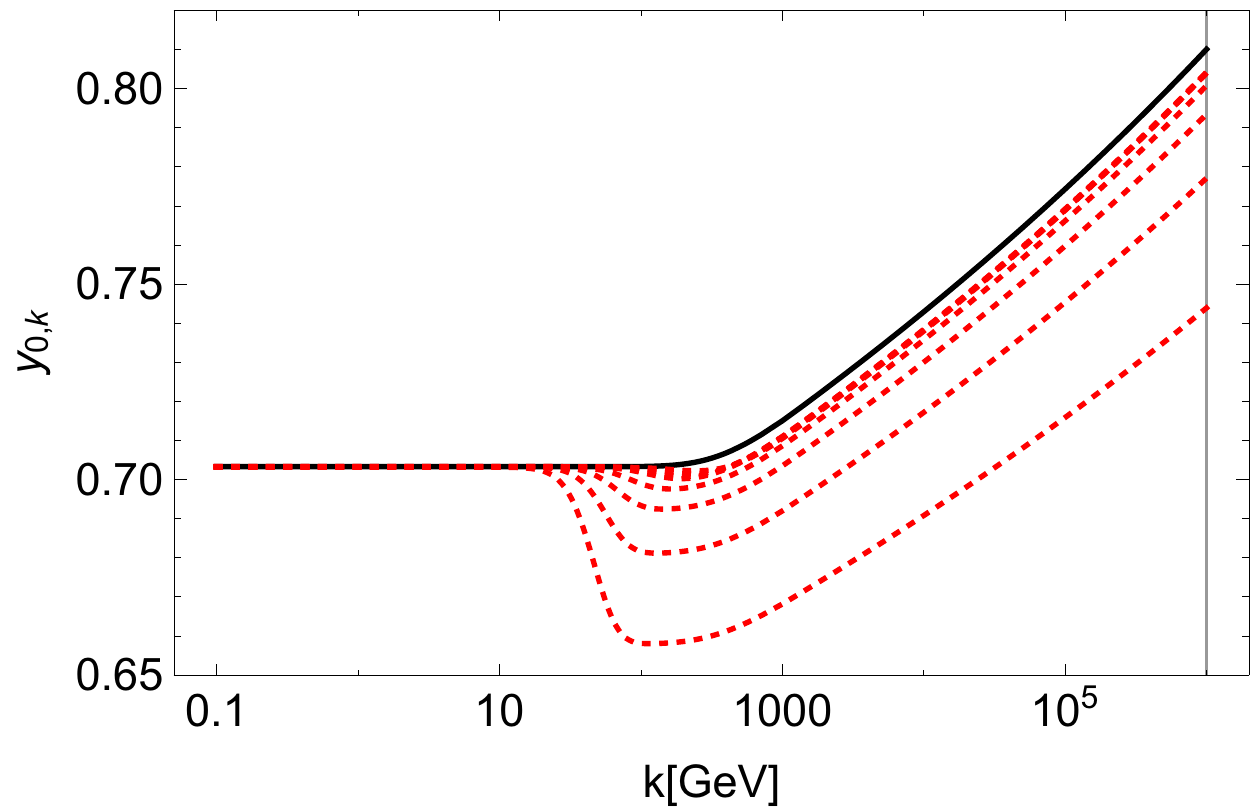}
\caption{Running of the Yukawa coupling $y_{0}$ for different truncation orders. The black solid lines depict the converged values within the first projection scheme. The red dotted lines show different orders $\Nh$ of the second projection scheme, see main text. The left panel shows the purely gauge-induced running Yukawa coupling. The middle and right panel show the flow of $y_{0}$ including also the scalar and fermionic interplay for the gauged and ungauged model.}
\label{fig:Running-y02}
\end{figure*}

Nevertheless, the modification of the flow due to the additional terms
contributes indirectly to the full flow as the decoupling of modes is
altered.  Within the first projection scheme, the quantum fluctuations
encoded in the threshold function $l_{1,1}^{(\rmF\rmB)}$ contribute
monotonically decresasing to the beta function. They approach zero
rapidly as soon as the flow approaches the decoupling region. As the
dimensionless expectation value $\kappa$ grows large ${\sim}\E^{-2t}$,
the flow of $y_0$ decouples as $\pt y_{0} \sim \kappa^{-3}$, since the
corresponding diagram has three internal lines given by two massive
top quarks and one massive scalar.

The other two contributions within the second projection scheme given on the right-hand side of Eq.~\eqref{eq:additional} show a different behavior. 
As they come with the opposite sign, they can lead to a decrease towards negative values of the beta function from the UV to IR for the ungauged model. For small $\kappa$, i.e., close to the transition from the SYM to the SSB regime, they only have a small impact on the beta function of $y_{0}$ as they contribute ${\sim}\kappa$. Deep in the IR where $\kappa$ grows large, their contribution is also suppressed according to $\kappa l_{2,1}^{(\rmF\rmB)} \sim \kappa^{-3}$ and $\kappa l_{1,2}^{(\rmF\rmB)} \sim \kappa^{-3}$. This can be seen diagrammatically in Fig.~\ref{fig:Feyn1} where the second and third diagram contain additional internal propagators due to the couplings to the condensate. These two diagrams correspond to the threshold functions $l_{2,1}^{(\rmF\rmB)}$ and $l_{1,2}^{(\rmF\rmB)}$, respectively. (Note, that $l_{n_1,n_2}^{(\rmF\rmB)} \sim \kappa^{-(n_1+n_2+1)}$ for $\kappa\gg 1$.) In the intermediate regime where $\kappa \sim 1$ the additional contribution to the beta function approaches a maximum which overwhelms the positive contribution from the standard $l_{1,1}^{(\rmF\rmB)}$ term near the region where the threshold effects set in. This leads to a negative beta function during the decoupling. Thus, the integrated running coupling $y_{0}$ develops a slight dip near the electroweak scale within the second projection scheme, visible in Fig.~\ref{fig:Running-y0}.

As the deep IR value of the Yukawa coupling $y_{0}$ is fixed by the measured top mass, the modification in the decoupling region leads to an altered running for the whole flow from the UV to the IR even though the flow equations are the same in the SYM regime. Only slight deviations between the different projections for $\Nh=0$ within the second projection scheme and $\Nh\geq1$ for the first projection scheme occur, because the whole flow is essentially governed by the Yukawa coupling $y_{0}$ itself. The small region in which the decoupling of modes is slightly altered mimics the contributions coming from higher-order operators in an appropriate manner in this model.

However, the situation is different for gauged Yukawa models. First of
all, the flow of $y_{0}$ is changed qualitatively by including the
strong force. The term $\sim g^{2}$ generated by the gauge boson
contributions in the flow equation for the Yukawa potential
\eqref{eq:flowYukawa} comes with a sign opposite to the fermion-scalar
induced part. Due to the size of the strong coupling constant, the
running Yukawa potential is mainly built up from this contribution and
the Yukawa coupling $y_{0}$ is increasing during the flow from the UV
towards the IR as can be seen in the right panel of
Fig.~\ref{fig:Running-y0}. This time, the contribution of $y_{1}$
relaxes the increase of $y_{0}$ as the gluon dominated RG flow
generates a small but negative value for this higher-dimensional
coupling which explains the fact that the black solid curve lies
slightly above the blue dashed curve for the gauged model.

The effective inclusion of higher-order couplings in the second projection rule exceeds their actual impact in the SSB regime. This can be traced back to the modified decoupling of modes. The strong coupling constant is already of order one close to the Fermi scale and drives the running Yukawa couplings. This leads to a much stronger modification of the decoupling turning the beta function to positive values. This causes the bulge in Fig.~\ref{fig:Running-y0} (right panel) which is more distinct than the dip within the ungauged model (left panel).

This is in agreement with the extended mean-field analysis: the RG flow originating from the second projection scheme can overrate the impact of the neglected higher-order Yukawa couplings which leads to larger Higgs masses in the gauged model.

\subsection{Convergence properties}
Let us finally investigate the convergence properties within this projection scheme by increasing $\Nh$. 
We perform this analysis first for the pure gauge induced running of the Yukawa sector, similar to the previous subsection. In the left panel of Fig.~\ref{fig:Running-y02} the running of $y_{0}$ is depicted again for different truncation orders for the gauged model with $\Lambda = 10^{6}$ GeV. The black solid curve shows the converged behavior within the first projection scheme. The red dotted curves show the flow given by the second projection scheme for $\Nh=0, \, 2, \, 5, \, 15,\, 30,\, 45$ from top to bottom. Indeed, we find a slow convergence towards the black solid line which can be seen by a depleting bulge which shifts to smaller scales. This convergence behavior for this particular approximation is in line with the extended mean-field analysis which suggests an $\Nh^{-1/2}$ behavior.

However, this slow convergence of the second projection rule is even spoiled as soon as we include also the fermion-scalar contribution to the flow. 
The middle and right panel show the convergence properties of the nonperturbative flow within our chosen truncation for the gauged and ungauged Yukawa model, respectively. 
The red dotted curves are computed in the second projection for $\Nh = 0,\, 1,\, 2,\, 3,\, 6,\, 9,\, 12,\, 15,\, 18$ from top to bottom for the gauged model. At first, we observe a convergence towards the result of the first projection scheme by increasing $\Nh$ which is again depicted as black solid line. 
We observe that the solutions of the differential equations drop below the converged result of the first projection rule for $\Nh \geq 10$ and depart from the converged solution.

This unexpected behavior is driven by the mutual interplay of the scalar and fermion fluctuations within the flow equation of the Yukawa coupling. 
Setting the argument coming from the scalar to zero in the threshold function on the right-hand side of the beta functional \eqref{eq:flowYukawa}, i.e., replacing $l_{1,1}^{(\rmF\rmB)d}(2 \trho h^2,u' + 2\trho u'' ; \eta_\psi,\eta_\phi)$ by $l_{1,1}^{(\rmF\rmB)d}(2 \trho h^2,0; \eta_\psi,\eta_\phi)$, leads to a convergence behavior similar to the one observed for the pure gauge induced part, see left panel in Fig.~\ref{fig:Running-y02}. 
By contrast, the scalar threshold effects and the extra terms coming from the second projection scheme spoil the convergence. 
There are two reasons for this: First of all, the decoupling regime is altered also in the gauged model within the second projection scheme which changes as $\Nh$ is varied, see Eq.~\eqref{eq:projectionGies4}. We get a second modification due to the different running of the pure scalar sector in addition.
As soon as the Yukawa coupling gets smaller during the flow, also the Higgs mass becomes smaller. Thus, the freeze out from the scalars sets in later while the top mass is fixed. Therefore, the scalar dominated fluctuations contribute for a longer RG time during the decoupling. The combination of these two effects, the increasing complex structure of the modified decoupling for large $\Nh$ as well as the smaller values of $\lambda_{2}$ during the decoupling, lead to a peak in the beta function in the SSB regime which causes the steep ascent in the running of $y_{0}$ towards the decoupling regime, see Fig.~\ref{fig:Running-y02} middle panel.

The same conclusions can be drawn for the ungauged model. For small
$\Nh$ the running of the Yukawa coupling stays close to the converged
values for the first projection rule. The red dotted lines depict the
solutions for $\Nh= 0,\, 1,\, 2,\, 3,\, 6,\, 9,\, 12,\, 15$ from top
to bottom in the right panel of Fig.~\ref{fig:Running-y02}. Due to the
smaller Yukawa couplings and thus smaller scalar quartic
self-interactions, the solutions also depart from the converged value
of the first projection scheme. They do not show any convergence within
the second projection scheme for $\Nh\gtrsim 10$ driven by the modified decoupling.

In summary, we conclude that the flow equations for the Yukawa couplings $y_{n}$ extracted from the first projection scheme should be preferred for the analysis of Higgs mass bounds in polynomial expansions. Even if the second scheme includes some parts of the truncated higher-order contributions, it shows worse convergence than the first projection rule. In the worst case, convergence can even get lost in polynomial expansions because of  the nontrivial interplay between the Higgs and Yukawa sector. Moreover, the additional contributions can exceed their actual impact as can be seen for small $\Nh$ for the gauged model for instance.

Our analysis finally resolves a puzzle observed in the literature:
within polynomial expansions using the second projection rule,
non-trivial fixed-point solutions of the RG flow have been found in
the simple Yukawa model \cite{Gies:2009hq} as well as a
Higgs-top-bottom chiral Yukawa model \cite{Gies:2014xha}, which could
serve as an indication that such models allow for a nontrivial UV
completion. Though these fixed points occurred either in unphysical
parameter regions or beyond the valididy range of the approximations,
the structure of the Yukawa flows in the SSB regime appeared to
provide evidence for the existence of a new fixed-point generating
mechanism by means of quasi-conformal condensates. By contrast,
fixed-point searches in the simple Yukawa model using numerical
shooting techniques going beyond the polynomial expansion did not find
any indication for such fixed points \cite{Vacca:2015nta}. From our
analysis, we now observe that the second projection scheme together
with the polynomial expansion indeed feature such fixed points,
whereas these solutions do not exist in the case of the first
projection scheme. Since the first projection scheme exhibits a much
better convergence behavior and is structurally more stable, we
conclude that the fixed-point solutions observed in the literature for
these systems are an artifact of the second projection scheme, leaving
the trivial Gau{\ss}ian fixed point as the only physical acceptable
fixed-point solution. We emphasize that our observations do not
exclude the possibilty of UV fixed-points generated from
quasi-conformal condensates in general. However, this mechanism
appears not to be active in the models studied so far.


\section{Impact of generalized Yukawa couplings on lower Higgs mass bounds}
\label{sec:pheno}

In order to extract the mass of the Higgs from the RG flows of the
scalar potential, we investigate the curvature of the Higgs effective
potential at the electroweak minimum in the deep IR as in the previous
section. We pay particular attention to the influence of generalized
Higgs-Yukawa interactions.  The corresponding RG flow equations are
provided in Sec.~\ref{sec:model}.

For our purpose, it suffices to study the flow of the coefficients in a polynomial approximation for the scalar potential Eq.~\eqref{eq:utrunc} and the Yukawa potential Eq.~\eqref{eq:htrunc}. We use a vanishing field amplitude $\kappa = 0$ in the SYM regime and a nonvanishing minimum $\kappa$ of the scalar potential in the SSB regime, $u'(\kappa)=0$. 
Inserting Eq.~\eqref{eq:utrunc} and Eq.~\eqref{eq:htrunc} into the Eqs.~\eqref{eq:flowPot}-\eqref{eq:anomalFermion}, we can extract the running of the various coefficients. We use the flow equations for the Yukawa couplings obtained from the first projection scheme in Sec.~\ref{sec:projection} leading to a fast convergence of our results.
We tested carefully that $\Nu=4$ and $\Nh=2$ is an optimal choice in terms of accuracy, precision, and computing time for the present models to investigate the mass spectrum.

Of course, the accuracy applies to the study of particle masses, i.e.,
the local properties of the potentials at the electroweak minimum. As
aforementioned, these local approximations of the potentials are not
sufficient to investigate global properties due to the limited radius
of convergence \cite{Gies:2013fua}. Thus, the issue of vacuum
metastability, in which a second minimum for the potential shows up,
is not fully accessible within this approximation. Still, the polynomial
flow is able to provide indications whether a second minimum may be
generated \cite{Borchardt:2016xju}.

For instance, in all cases studied so far where the polynomial flow is stable for all $k$, also the full potential flow does not exhibit a second minimum. In case the polynomial flow develops a second minimum besides the Fermi minimum at some RG scale, also the full potential flow acquires a second minimum which survives the RG flow rendering the effective Higgs potential metastable \cite{Borchardt:2016xju}. In the following, we use this circumstantial evidence to classify the stability properties of the model using the polynomial flows.

Assuming that the RG flow is only dominated by perturbatively renormalizable operators, the stability issue is related to the fact that the quartic Higgs self-coupling vanishes at the high momentum cutoff scale of the theory $\lambda_{2,\Lambda} = 0$. The consistency condition to have a well defined partition function, i.e., a scalar potential which is bounded from below, defines the lower mass bound within quartic bare potentials, $\lambda_{2,\Lambda} \geq 0$. Increasing the bare quartic coupling results in larger Higgs masses. By contrast, violating the consistency condition would in principle lead to a smaller Higgs mass but at the same time an ill-defined quantum field theory.

Beyond the class of bare $\phi^4$ potentials, the regularized field theory might be stabilized by higher-dimensional perturbatively nonrenormalizable operators such that the scalar potential exhibits only one minimum even if the quartic coupling is negative \cite{Gies:2013fua,Gies:2014xha,Eichhorn:2015kea,Hung:1995in,Casas:2000mn,Burgess:2001tj}. 
Already the simplest possible extension of the bare potential by a $\lambda_{3,\Lambda} \phi^{6}$ term can fulfill the requirements. Generically, the presence of the additional RG irrelevant operator modifies the running of the quartic coupling only slightly. Thus, $\lambda_{2}$ still drops below zero at approximately the same UV scale as in the $\phi^4$ case. But the form of the UV potential can be modified by a positive bare $\lambda_{3}$ coupling such that the potential is still stable even if this irrelevant coupling dies out during the RG flow. 
Assuming that the dimensionless higher-order coupling
$\lambda_{3}$ is of order one at the standard model cutoff scale, the
lower consistency mass bound for the Higgs can be shifted by $1$-$2$
orders of magnitude towards larger cutoff scales, i.e.,
$m_{\text{H}}(\Lambda)\to m_{\text{H}}(\Lambda')$, where $\Lambda'\simeq
\mathcal{O}(10 \dots 100)\times \Lambda$. A further increase of the
cutoff scale, i.e., diminishing of the lower Higgs mass bound, seems
not possible with this simple modification of the bare action without
generating a metastable potential \cite{Borchardt:2016xju}.

Let us now study the influence of a generalized bare Yukawa sector of
the theory, as the Yukawa coupling is the dominant quantity regarding
the lower Higgs mass bound. The convergence tests of the preceding
section used $y_{1,\Lambda}=y_{2,\Lambda}=\cdots = 0$ as initial
conditions.

The situation becomes more interesting if we already allow for a generalized Yukawa potential in the bare action $h_\Lambda(\trho)$ and investigate its influence on the conventional lower Higgs mass bound obtained for $h_\Lambda(\trho)=y_{0,\Lambda}$, and $\lL=\lambda_{3,\Lambda}=\cdots = 0$. Indeed, we find Higgs masses below the lower bound by the rather mild modification of choosing $y_{1,\Lambda}>0$. The size of this shift is roughly of the same order of magnitude as already observed by allowing for generalized scalar bare potentials $u_\Lambda$ \cite{Gies:2013fua,Gies:2014xha,Borchardt:2016xju,Chu:2015nha}.  
This confirms our expectation that  a modification of the running of the Yukawa sector $h$, which builds up the lower Higgs mass bound, could lead to such a reduction. Comparing the resulting flows of $y_0$ for $y_{1,\Lambda}=0$ and $y_{1,\Lambda}>0$ shows an insignificant shift for the values of $y_0(k)$ such that this mechanism is not able to explain the observed shift in the Higgs mass. 
 
The modification of the bare Yukawa functional also influences the flow of the scalar potential in a direct manner. This effect can already be investigated on the mean-field level. For this, we slightly rewrite the mean-field effective scalar potential,
 \begin{align*}
  U^{\text{MF}}(\rho) &= U_{\Lambda}(\rho) - \frac{\big[H_\Lambda(0)\big]^2 \Nc \Lambda^2\rho}{8\pi^2} \\
  &\quad + \frac{\Nc}{8\pi^2}  \Bigg[ - \big[H_\Lambda(\rho)-H_\Lambda(0)\big]^2 \Lambda^2\rho  \\
  &\qquad\qquad +  2\big[H_\Lambda(\rho)\big]^4 \rho^2 \ln{\left(1 + \frac{\Lambda^2}{\big[H_\Lambda(\rho)\big]^2\rho}\right)} 
  \Bigg],
 \end{align*}
where we have separated the interaction part of the Fermion determinant in the second and third line. This is a positive monotonically increasing function for $H_\Lambda(\rho) = y_{0,\Lambda}$. Thus, no metastability can be induced for vanishing generalized Yukawa interactions at the cutoff scale if the bare scalar potential is of quartic type \cite{Gies:2014xha}. Obviously, this function can lose its positivity property if we allow for generalized Yukawa interactions beyond $H_\Lambda = y_{0,\Lambda}$, depending on the precise shape of the in principle arbitrary $H_\Lambda(\rho)$. 
Hence, generalized bare Yukawa potentials can affect the scalar potential in two ways. First, smaller Higgs masses than the conventional lower bound can be obtained. In addition, it is also possible that the fermion interaction can induce a metastability even in the class of $\phi^4$ bare potentials depending on how the positivity property is circumvented.

Also, this fact can be seen from the flow equation of
$\lambda_2$. While the usual Yukawa coupling contributes according to
$-y_0^4$ and thus leads to an increase of the quartic coupling towards
the IR, an additional contribution coming from $y_1$ contributes
according to $+y_1 y_0$. Thus a sufficiently large value for
$y_{1,\Lambda}$ can lead to a decrease of $\lambda_2$ near the cutoff
scale.  Starting with a vanishing interacting scalar bare potential
($\lL=\lambda_{3,\Lambda}=\cdots=0$), $\lambda_2$ runs to negative
values while $\lambda_3$ becomes positive (and similar for the higher-order scalar couplings) driven by $y_1$. In the course of the flow,
$y_1$ quickly becomes small as expected from power counting in the
vicinity of the Gau\ss{}ian fixed-point. Then, the flow equations
become dominated by the pure Yukawa interaction term $\sim
-y_0^{2n}$. Roughly speaking, we flow into the class of generalized
bare potentials with $\lL<0$ and $\lambda_{3,\Lambda}>0$ after a short
RG time. In complete analogy to the generalized bare potentials with
$\lL<0$ and stabilized by $\lambda_{3,\Lambda}$, a metastability
occurs during the RG flow if $y_{1,\Lambda}$ exceeds a critical value,
driving $\lambda_2$ towards too small values.

\begin{figure*}
  \centering
 \includegraphics[width=0.95\columnwidth]{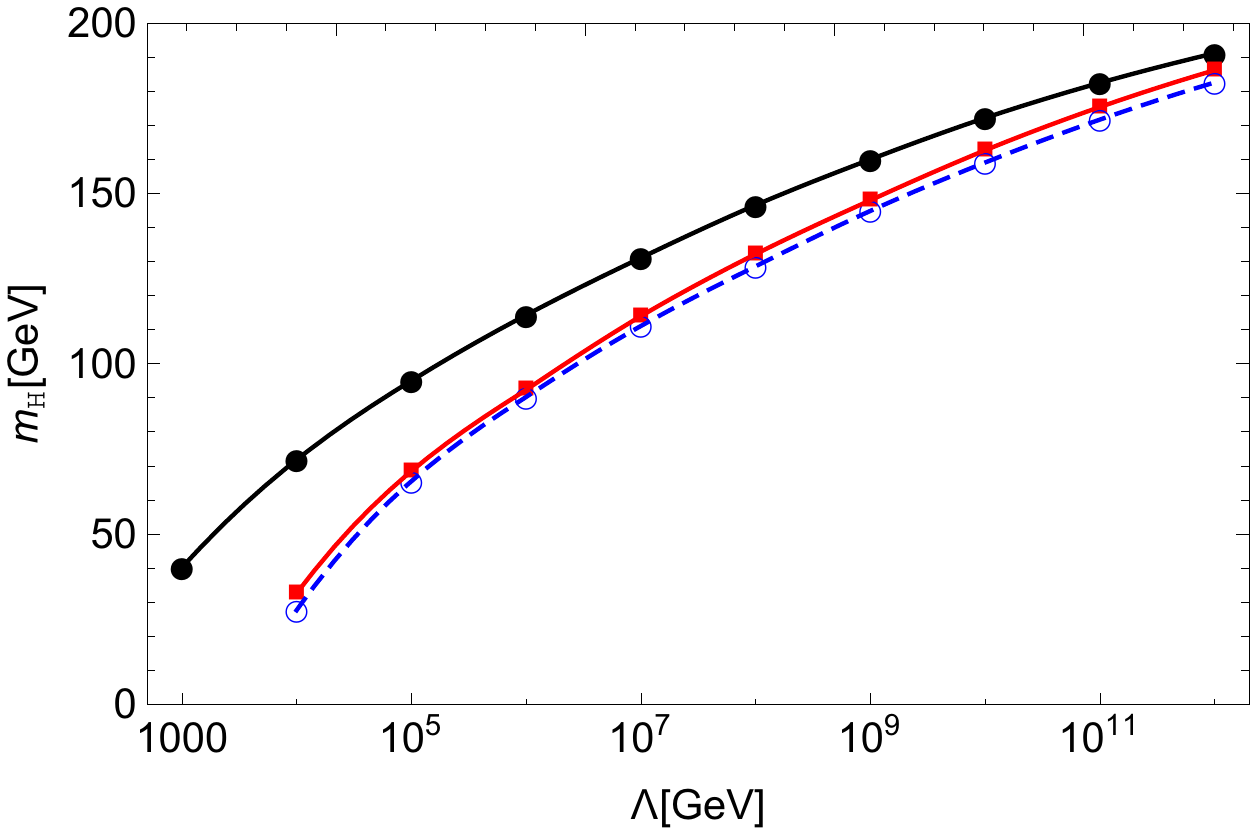}
 \hfill
 \includegraphics[width=0.95\columnwidth]{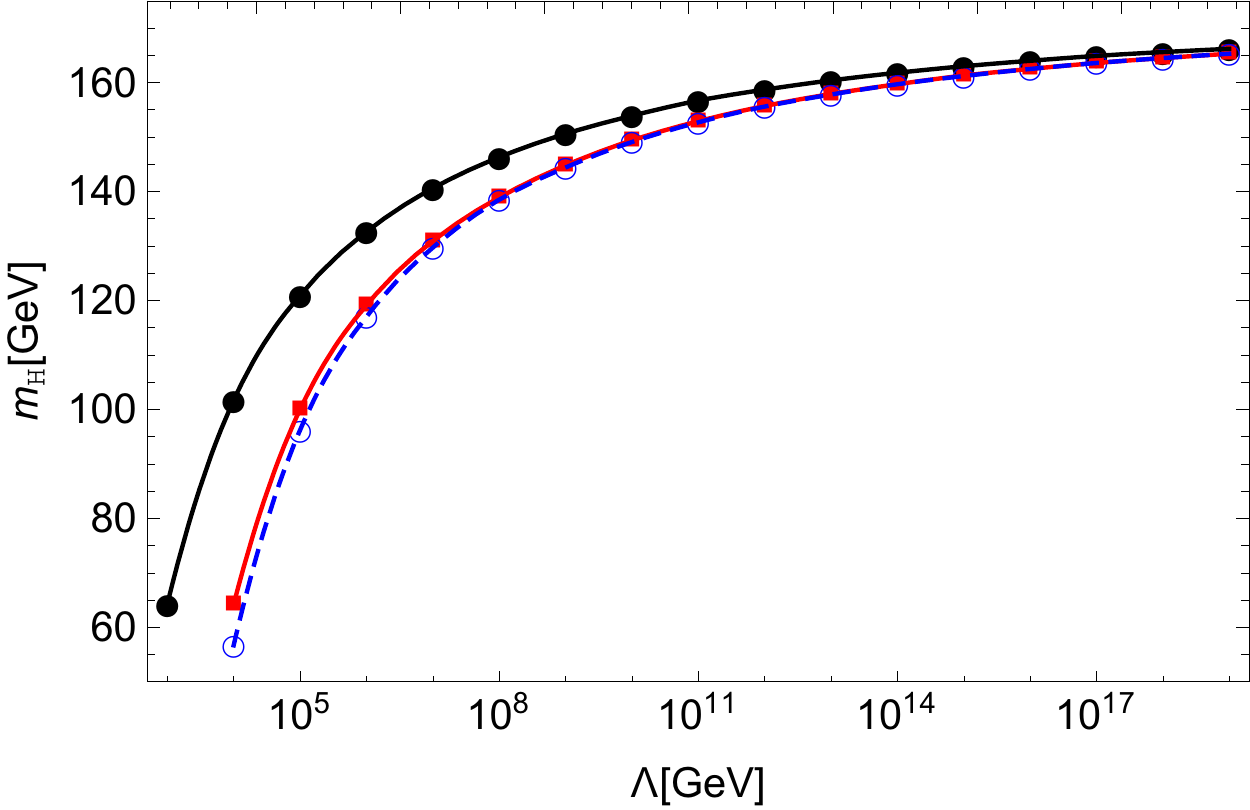}
 \caption{Higgs mass versus cutoff. The black curve denotes the conventional lower Higgs mass bound obtained from a theory which only include perturbatively renormalizable operators at the UV cutoff scale. The red curve gives a lower bound in the class of generalized scalar bare potentials of $\phi^6$ type with the requirement that no metastability occurs during the RG flow. The Higgs masses given by the blue dashed line are obtained from a further generalization to Yukawa potentials of the form $y_{0,\Lambda}+y_{1,\Lambda}\trho$. Left panel: ungauged model. Right panel: gauged model. The relative shift of the Higgs mass bound is $0.9$ GeV at the Planck scale for the gauged model.}
 \label{fig3:BelowHiggsMassh1}
\end{figure*}
 
Following the same strategy as in the case of generalized bare potentials, we trust the polynomial expansion as long as no second minimum deeper than the minimum at the electroweak scale emerges during the RG flow \cite{Borchardt:2016xju}. Fixing $\lL=\lambda_{3,\Lambda}=\cdots=0$ and increasing $y_{1,\Lambda}$ until a potential second minimum becomes the global one, we obtain Higgs masses which we already covered in the class of generalized bare potentials. The naive expectation, that a combination of both mechanisms (generalized $U_\Lambda$ and $H_\Lambda$) could lead to a further decrease of the lower Higgs mass bound, is investigated in the following. 

The basic idea is, that the vanishing of $\lambda_3$ can be delayed for a small RG time by choosing, for instance, $\lambda_{3\Lambda}=3$ in addition to $y_{1,\Lambda}>0$. Hence, lower values for $\lambda_2$ during the flow are accessible and the Higgs mass decreases. 
 However, this further diminishing is rather small, because the running of the higher-order couplings is strongly affected by their infrared attractive pseudo-fixed points which cannot be circumvented by this strategy. In order to illustrate this fact, we depict the lowest accessible Higgs masses from the different generalizations of the bare action in Fig.~\ref{fig3:BelowHiggsMassh1} for the ungauged (left panel) as well as the gauged model (right panel). We compare the Higgs masses obtained for three different scenarios: (A) for the conventional lower Higgs mass bound ($\lL=\lambda_{3,\Lambda}=\cdots=0$ and $y_{1,\Lambda}=y_{2,\Lambda}=\cdots=0$), (B) for the lower bound obtained from the class of $\phi^6$ bare potentials with the requirement of stable scalar potentials during the entire RG flow, and (C) a similar lower bound but for the parameters $\lL=0$, $\lambda_{3,\Lambda}=3$, and the value of $y_{1,\Lambda}$ put to the transition where the scalar effective potential becomes metastable. We emphasize again, that the present toy models should not be used for a direct quantitative comparison with standard-model parameters, such as the mass of the Higgs boson of $\mH = 125$ GeV. This is mainly because of a different particle content, but also due to differences in the regularization scheme. Nevertheless, we expect that these differences partly cancel if relative quantities are studied, such as the relative change of the lower Higgs mass bound.
For the gauged model, the relative shift of the Higgs mass bound is at least $0.9$ GeV at the Planck scale for both described scenarios and increases for smaller cutoff values.

 In addition, we have tested combinations including nonvanishing bare
 parameters for $\lambda_{2,\Lambda}$, $\lambda_{3,\Lambda}$,
 $\lambda_{4,\Lambda}$, $y_{1,\Lambda}$, and $y_{2,\Lambda}$ for
 selected cutoff values. No significant diminishing compared to
 previous investigations has been found, as expected from
 power-counting.  Thus, generalized Yukawa interactions can indeed
 influence the running of the scalar sector such that the Higgs mass
 consistency bound can be diminished but their impact is of the same
 level as a modification of the bare Higgs potential. As the system in
 $d=4$ is dominated by the Gau{\ss}ian fixed point, probably any
 polynomial interaction term which could lead to a lowering of this
 bound will only shift by the same amount of scales -- at least as
 long as it has or generates sufficient overlap with dimension 6
 operators. From our studies, we expect that various combinations of
 these effects will not lead to any further significant change of the
 consistency bound, as long as the flow is initiated in the region
 dominated by the Gau\ss{}ian fixed point.

 Recently, another interesting mechanism was proposed to lower the
 Higgs mass consistency bound. In \cite{Jakovac:2015kka}, higher-order
 operators were imposed to drive the slope of the quartic coupling
 to zero in addition to a vanishing $\lambda_{2}$ at the UV cutoff
 $\Lambda$, rather than to stabilize the potential with a negative
 $\lambda_{2}$ at $\Lambda$. Such a scenario can be built on
 generalized bare Yukawa couplings, for instance, by the requirement that $h_{1,\Lambda} = h_{0,\Lambda}^{3}$ and vanishing other
 higher-order couplings. Then, the flow equations can be integrated
 upwards avoiding a negative $\lambda_{2}$. As long as a few terms of
 a polynomial truncation of the scalar potential are considered, the
 flow seems to be stable. This leads to an effective higher cutoff
 $\Lambda' = \Lambda$ and thus to a diminishing of the lower
 bound.

 Though the scenario is attractive and appears to be viable, we
 have not been able to construct a robust example. Within polynomial
 expansions, this is because truncated polynomial potentials
 unfortunately lose their reliability after a few scales for two
 reasons. First of all, the neglected higher-order operators grow
 large such that it becomes questionable whether a finite-order
 approximation can reliably track the relevant structure of the scalar
 potential. This calls at least for suitable resummations in this
 region. Second, within our analyses, nontrivial minima generically
 show up during the flow rendering the potential metastable. Previous
 studies indicate that these metastabilities survive the RG running of
 the full potential, even if they seem absent in the ($k\to0$)
 effective potential in a polynomial Taylor series
 \cite{Borchardt:2016xju}. In summary, we observe that a further
 significant diminishing of the consistency bound is not possible, as
 long as the flow is close to the Gau{\ss}ian fixed point, and thus in
 a weakly coupled controllable region beyond the class of
 perturbatively renormalizable operators.

\section{conclusion}
\label{sec:conclusions}
In this work, we have investigated the impact of a generalized Yukawa function $H(\rho)$ rather than a simple coupling on the lower Higgs mass consistency bound from a functional RG perspective. 

As a technical though important step in practice, we have first
analyzed different projection schemes on the higher-order Yukawa
couplings that have been used in the literature. While both
projections lead to the same flow equation for the Yukawa potential,
we observe differences regarding the convergence properties as soon as
the systems are truncated to finitely many couplings. The flow
equations extracted from the projection rule given by
Eq.~\eqref{eq:projectionPawlowski} show fast convergence properties
for the application discussed in this work. By contrast, the flow
equations based on the second projection rule
Eq.~\eqref{eq:projectionGies} lead to a slower convergence or even a
loss of convergence for local expansions. This is unexpected as the
second projection scheme includes part of the truncated information of the neglected
higher-order couplings. It turns out that these additional
contributions can overestimate their true impact resulting in a
non-appropriate decoupling of the massive modes, spoiling
convergence of local expansions.

Based on the fast converging projection, we have then investigated the
impact of higher-dimensional bare Yukawa couplings on the lower Higgs
mass bound. Choosing a nonvanishing value for the bare coupling of the
dimension-six operator $y_{1,\Lambda}$, it is possible to lower the
conventional mass bound by about a similar amount as was found earlier
upon the inclusion of a $\sim\lambda_{3,\Lambda}\phi^6$ operator
\cite{Gies:2013fua,Gies:2014xha,Eichhorn:2015kea,Borchardt:2016xju,Chu:2015nha}.
This is another example for the fact that the consistency bound
linking a consistent quantum field theory defined with a cutoff
relaxes the conventional lower mass bound.

The inclusion of both dimension-six operators in the bare action
relaxes the bound even further, though the combined effect is small
compared to the relaxation of the bound by each single dimension-six
operator. The fact that the shift of the bound is non-additive in both
operators is a consequence of the power-counting RG flow in the
weak-coupling region, inducing a simultaneous power-law depletion of
higher-dimensional operators. We emphasize that the result of an
initially strongly coupled bare action on the consistency bounds still
remains an interesting open problem both for the sector of the
generalized Yukawa couplings as well as for the purely scalar sector.

\section*{Acknowledgments}
We thank Julia Borchardt, Tobias Hellwig, Istvan Nandori, Jan Pawlowski, Manuel Reichert, Fabian Rennecke, and Luca Zambelli for valuable discussions. We acknowledge support by the DFG under Grants No.  GRK1523/2 and No. Gi328/7-1. RS thanks the Carl-Zeiss Stiftung for support through a postdoc fellowship.

\appendix

\section{Threshold functions}
\label{appA}

The threshold functions appearing in the flow equations depend on the chosen IR
regulator which parametrizes the details of the momentum-shell
integration. Using the abbreviations
\begin{align*}
P_\rmB(q) = q^2 (1+r_{\rmB,k}(q)), \quad P_\rmF(q) = q^2(1+r_{\rmF,k}(q))^2,
\end{align*}
with $r_{\rmB,k}$ and $r_{\rmF,k}$ the regulator shape functions for the bosonic and fermionic degrees of freedom respectively, 
as well as
\begin{align*}
\tilde{\partial}_t = \sum_{i=F,B} \int_q \frac{\partial_t (Z_{i,k}\, r_{k,i}(q))}{Z_{i,k}} \frac{\delta}{\delta r_{k,i}(q)},
\end{align*}
the threshold functions are defined as
\begin{widetext}
\begin{align*}
l_n^{(\rmB)d}[\omega;\eta_\rmB] &= \frac{n+\delta_{n,0}}{4v_d} k^{2n-d} \int_q \left[ \frac{\pt R_{\rmB,k}(q)}{Z_{\rmB,k}} (P_\rmB(q)+\omega k^2)^{-(n+1)}\right],\\
l_n^{(\rmF)d}[\omega;\eta_\rmF] &= \frac{n+\delta_{n,0}}{2v_d} k^{2n-d} \int_q \left[ \frac{P_\rmF(q)\,\pt (Z_{\rmF,k} \, r_{\rmF,k}(q))}{Z_{\rmF,k}(1+r_{\rmF,k})}(P_\rmF(q)+\omega k^2)^{-(n+1)}\right],\\
l_{n_1,n_2}^{(\rmF\rmB)d}[\omega_1,\omega_2;\eta_\rmF,\eta_\rmB] &= -\frac{1}{4v_d}k^{2(n_1+n_2)-d} \int_q \tilde{\partial}_t \left[ \frac{1}{(P_\rmF(q)+k^2 \omega_1)^{n_1}(P_\rmB(q)+k^2 \omega_2)^{n_2}}\right],\\
m_4^{(\rmF)d}[\omega;\eta_\rmF] &= -\frac{1}{4v_d} k^{4-d} \int_q q^4 \tilde{\partial}_t \left[\frac{\partial}{\partial q^2}\frac{1+r_{\rmF,k}(q)}{P_\rmF(q)+k^2 \omega}\right]^2,\\
m_2^{(\rmF)d}[\omega;\eta_\rmF] &= -\frac{1}{4v_d} k^{6-d} \int_q q^2 \tilde{\partial}_t \left[\frac{\left(\frac{\partial}{\partial q^2}P_\rmF(q)\right)}{P_\rmF(q)+k^2 \omega}\right]^2,\\
m_{2}^{(\rmB)d}[\omega;\eta_\rmB] &= -\frac{1}{4v_d} k^{2(n_1+n_2-1)-d} \int_q q^2 \tilde{\partial}_t \left[ \frac{\left(\frac{\partial}{\partial q^2}P_\rmB(q)\right)}{P_\rmB(q)+k^2 \omega} \right]^{2},\\
m_{n_1,n_2}^{(\rmF\rmB)d}[\omega_1,\omega_2;\eta_\rmF,\eta_\rmB] &= -\frac{1}{4v_d} k^{2(n_1+n_2-1)-d} \int_q q^2 \tilde{\partial}_t \left[ \frac{1+r_{\rmF,k}(q)}{(P_\rmF(q)+k^2 \omega_1)^{n_1}} \frac{\left(\frac{\partial}{\partial q^2}P_\rmB(q)\right)}{(P_\rmB(q)+k^2 \omega_2)^{n_2}} \right],\\
\tilde{m}_{n_1,n_2}^{(\rmF\rmB)d}[\omega_1,\omega_2;\eta_\rmF,\eta_\rmB] &= -\frac{1}{4v_d}k^{2(n_1+n_2)-d}\int_q \tilde{\partial}_t \left[ \frac{1+r_{\rmF,k}(q)}{(P_\rmF(q)+k^2 \omega_1)^{n_1}} \frac{1}{(P_\rmB(q)+k^2 \omega_2)^{n_2}} \right].
\end{align*}
\end{widetext}
Throughout our analysis we use the Litim regulator which is optimized for the derivative expansion \cite{Litim:2001up}:
\begin{align*}
r_{\rmB,k}(q)=\left( \frac{k^2}{q^2}-1\right) \Theta(k^2 - p^2).
\end{align*}
The fermion shape function is chosen such that $(1+r_{\rmF,k}(q))^2=(1+r_{\rmB,k}(q))$. 
The momentum integral of the threshold functions can be solved analytically for the Litim regulator. They read
\begin{widetext}
\begin{align*}
l_n^{(\rmB)d}[\omega;\eta_\rmB] &= \frac{2(n+\delta_{n,0})}{d}\left(1-\frac{\eta_{\rmB,k}}{d+2}\right)\frac{1}{(1+\omega)^{n+1}},\\
l_n^{(\rmF)d}[\omega;\eta_\rmF] &= \frac{2(n+\delta_{n,0})}{d}\left(1-\frac{\eta_{\rmF,k}}{d+1}\right)\frac{1}{(1+\omega)^{n+1}},\\
l_{n_1,n_2}^{(\rmF\rmB)d}[\omega_1,\omega_2;\eta_\rmF,\eta_\rmB] &= \frac{2}{d}\frac{1}{(1+\omega_1)^{n_1}(1+\omega_2)^{n_2}}\left[\frac{n_1}{1+\omega_1} \left(1-\frac{\eta_{\rmF,k}}{d+1}\right) + \frac{n_2}{1+\omega_2} \left(1-\frac{\eta_{\rmB,k}}{d+2}\right)\right],\\
m_4^{(\rmF)d}[\omega;\eta_\rmF] &= \frac{1}{(1+\omega)^4}+\frac{1-\eta_{\rmF,k}}{d-2}\frac{1}{(1+\omega)^3}-\left(\frac{1-\eta_{\rmF,k}}{2d-4}+\frac{1}{4}\right)\frac{1}{(1+\omega)^2},\\
m_2^{(\rmF)d}[\omega;\eta_\rmF] &= \frac{1}{(1+\omega)^4},\\
m_{2}^{(\rmB)d}[\omega;\eta_\rmB] &= \frac{1}{(1+\omega_1)^4},\\
m_{n_1,n_2}^{(\rmF\rmB)d}[\omega_1,\omega_2;\eta_\rmF,\eta_\rmB] &= \left(1-\frac{\eta_{\rmB,k}}{d+1}\right)\frac{1}{(1+\omega_1)^{n_1}(1+\omega_2)^{n_2}},\\
\tilde{m}_{n_1,n_2}^{(\rmF\rmB)d}[\omega_1,\omega_2;\eta_\rmF,\eta_\rmB] &= \frac{2n_2}{(d-1)}\left(1-\frac{\eta_{\rmB,k}}{d+1}\right)\frac{1}{(1+\omega_1)^{n_1}(1+\omega_2)^{n_2+1}},\\
&\quad -\frac{1+\omega_1-2n_1}{d-1}\left(1-\frac{\eta_{\rmF,k}}{d}\right)\frac{1}{(1+\omega_1)^{n_1+1}(1+\omega_2)^{n_2}}.
\end{align*}
\end{widetext}

\bibliography{bibliography}

\end{document}